\documentclass[pra,reprint,groupedaddress]{revtex4-2}

\pdfoutput=1

\usepackage{amsmath}
\usepackage{amsfonts}
\usepackage{graphicx}
\usepackage{epstopdf}
\usepackage{newtxtext,newtxmath}
\usepackage{comment}
\usepackage{dcolumn}
\usepackage{natbib}
\usepackage{lipsum}
\usepackage{chemfig}

\usepackage{hyperref}
\hypersetup{
    colorlinks=true,       
    linkcolor=blue,          
    citecolor=blue,        
    filecolor=magenta,      
    urlcolor=blue           
}

\newcommand{\Eb}{\varepsilon _b}
\newcommand{\Eres}{\varepsilon_\nu}
\newcommand{\Ebar}{\bar{\varepsilon}}

\newcommand{\Gn}{\Gamma_\nu}
\newcommand{\Gne}{\Gamma_\nu^{e}}

\newcommand{\Bn}{\beta_\nu}

\newcommand{\Zeffm}{Z_\text{eff}^{(m)}}

\newcolumntype{d}[1]{D{.}{.}{#1}}

\let\originalleft\left
\let\originalright\right
\renewcommand{\left}{\mathopen{}\mathclose\bgroup\originalleft}
\renewcommand{\right}{\aftergroup\egroup\originalright}

\newcommand{\Zeff}{Z_\text{eff}}

\newcommand{\negphantom}[1]{\ifmmode\settowidth{\dimen0}{$#1$}\else\settowidth{\dimen0}{#1}\fi\hspace*{-\dimen0}}

\begin{document}

\frenchspacing

\title{Effect of chlorination on positron binding to hydrocarbons: experiment and theory}
\author{A. R. Swann}\email{a.swann@qub.ac.uk}
\author{G. F. Gribakin}\email{g.gribakin@qub.ac.uk}
\affiliation{School of Mathematics and Physics, Queen's University Belfast, University Road, Belfast BT7 1NN, United Kingdom}
\author{J. R. Danielson}\email{jdan@physics.ucsd.edu}
\author{S. Ghosh}\email{soumen@physics.ucsd.edu}
\author{M. R. Natisin}\altaffiliation{Present address: Air Force Research Laboratory, Aerospace Systems Directorate, Rocket Propulsion Division, Jacobs Technology Inc, Edwards AFB, CA 93524, USA}
\author{C. M. Surko}\email{csurko@ucsd.edu}
\affiliation{Department of Physics, University of California San Diego, La Jolla, California 92093, USA}
\date{\today}

\begin{abstract}
Measured and calculated positron binding energies are presented for a range of hydrocarbons with up to six carbon atoms (viz., methane, acetylene, ethylene, ethane, propane, butane, and hexane) and their chlorinated counterparts. Both experiment and theory confirm the large effect that the chlorine atoms have on the positron binding energy and the strong sensitivity of the binding energy to the exact position of the chlorine atoms. 
The experimental binding energies have been obtained by measuring positron resonant annihilation using a trap-based positron beam. The calculations are performed using the previously developed model-correlation-potential method [A.~R. Swann and G.~F. Gribakin, \href{https://doi.org/10.1063/1.5055724}{J. Chem. Phys. \textbf{149}, 244305 (2018)}]. The overall trends are discussed with regard to the molecular polarizability, dipole moment, and geometry. Good agreement between theory and experiment is found, with the exception of the chlorinated ethylenes and chlorinated hexane. Calculations of the electron-positron annihilation rate in the bound state are also presented.
\end{abstract}

\maketitle

\section{\label{sec:intro}Introduction}

The positron ($e^+$) is a useful tool in many areas of science, e.g.,  in fundamental tests of QED and the standard model \cite{Karshenboim05,Ishida14,ALEPH06}, astrophysics \cite{Guessoum14}, condensed-matter physics \cite{Tuomisto13}, and medicine \cite{Wahl02}. Despite this, there remains much about interactions between positrons and matter that is not well understood. One such topic is positron binding to atoms and molecules.

The existence of positron-atom bound states was predicted by many-body-theory calculations in 1995 \cite{Dzuba95}, and two years later, variational calculations confirmed that a positron can bind to lithium  \cite{Ryzhikh97,Strasburger98}. A wealth of calculations for other atoms followed; see Refs.~\cite{Mitroy02,Dzuba12,Harabati14} and references therein.
However, the existence of such bound states has not been verified experimentally, chiefly due to difficulties with obtaining the atomic species in the gas phase and devising an unambiguous detection scheme (although a number of proposals have been made \cite{Mitroy99,Dzuba10,Surko12,Swann16}).

On the other hand, for polyatomic molecules, calculations of positron binding have proven to be challenging. This is due to the strong electron-positron correlations that determine the binding energy and, for nonpolar species, ensure the very existence of the bound state. Most calculations have been for strongly polar molecules, where the existence of a bound state is  guaranteed even at the static, Hartree-Fock level of approximation
\footnote{A molecule with a dipole moment greater than 1.625~D is guaranteed to have a bound state for a positron (or an additional electron) \cite{Crawford67}, although this critical dipole moment increases if the molecule is rotating \cite{Garrett71}.},
although the calculated binding energy increases greatly when electron-positron correlations are included \cite{Tachikawa11,Tachikawa12,Tachikawa14}. \textit{Ab initio} calculations have so far failed to predict binding to nonpolar species.
On the other hand, positron-molecule binding energies have been determined experimentally for around 90 species using a trap-based positron beam \cite{Gilbert02,Barnes03,Barnes06,Young07,Young08_large,Young08_small,Danielson09,Danielson10,Danielson12,Danielson12a,Jones13,Natisin_thesis}.
These experiments measure the binding energies by observing vibrational Feshbach resonances (VFRs) in the positron-energy dependence of the annihilation rate (i.e., annihilation spectra).
The key point here is that for molecules capable of binding the positron, two mechanisms contribute to the annihilation process \cite{Gribakin00,Gribakin01}: in addition to the possibility of the positron annihilating with a molecular electron ``in flight'' (\textit{direct} annihilation), the positron can also become attached to the molecule by transferring the excess energy into molecular vibrations and forming a quasibound state, known as the VFR. The positron capture increases the dwell time of the positron near the molecule \footnote{In addition to the possibility of the bound positron annihilating with a molecular electron, a VFR can also decay by detachment of the positron, accompanied by deexcitation of the molecular vibrational state.}, resulting in strong enhancements of the annihilation rate at specific positron energies (\textit{resonant} annihilation). These energies depend on the positron binding energy $\varepsilon_b$ and the vibrational excitation energy of the molecule. For the excitation of a vibrational mode $\nu$ with frequency $\omega_\nu$, the energy of the VFR is
\begin{equation}\label{eq:eps_nu}
\varepsilon_\nu = \hbar \omega_\nu - \varepsilon_b ,
\end{equation}
which allows one to determine the binding energy from the positions of the resonances
\cite{Gilbert02,Gribakin10}. In contrast to the calculations of positron-molecule binding energies, most of the measurements  have been for nonpolar or weakly polar species. 

In this work we explore positron binding to chlorinated hydrocarbon molecules.
Measured binding energies have previously been reported for a range of singly and multiply chlorinated methanes and ethylenes \cite{Young08_small,Jones13,Natisin17,Gribakin17}, and $n$-hexyl chloride \cite{Young07}. Here we summarize these data and report newly measured binding energies for methylene chloride, \textit{cis}-1,2-dichloroethylene, trichloroethylene, and singly and doubly chlorinated propanes. Besides these chlorine-containing species, we also present improved measurements for propane, $n$-butane, and $n$-hexane (earlier measurements for these molecules were described in Ref.~\cite{Young08_large}), and a new measurement for isobutane.

Positron-molecule binding energies were obtained from analysis of the resonances observed in annihilation spectra as a function of incident positron energy. This requires knowledge of the vibrational modes of the molecule and is a sensitive function of the positron beam parameters. Presented here are examples of annihilation spectra and binding-energy data for a range of molecules, with emphasis on the chlorinated molecules that are the focus of this paper. These data were taken over many years, from the beginning of the binding-energy measurements in 2002 \cite{Gilbert02} to the present, using positron beams with different parameters and different analysis procedures. The new and previous data sets are corrected for the differences in beam parameters, and the binding energies are obtained from a consistent analysis of the spectra for these molecules.

To analyze and understand the observed trends, we carry out calculations of the binding energies using the model-correlation-potential method \cite{Mitroy02a,Swann18}. This is not an \textit{ab initio} approach. However, it provides a good, effective description of positron-atom interactions \cite{Mitroy02a}, working well for both small molecules, such as hydrogen cyanide \cite{Swann18}, and large polyatomics, namely alkanes \cite{Swann19,Swann20a}, where it has provided a number of useful insights and an accurate description of the experimental data. This method has also been applied to positron elastic scattering and direct annihilation in small molecules \cite{Swann20}. The model contains one free parameter for each type of atom within the molecule, which can be fixed by reference to accurate calculations or experimental data for binding or scattering.
In addition to the species with measured binding energies, we have performed calculations for chlorinated acetylenes, ethanes, and butanes.

The paper is structured as follows. In Sec.~\ref{sec:Method} we describe the methodology of the measurements and calculations of the positron binding energies. The results are presented in Sec.~\ref{sec:results}. We conclude in Sec.~\ref{sec:conc} with a summary and outlook.


\section{Methodology}\label{sec:Method}

\subsection{\label{sec:exp}Measurements}

Positron annihilation rates are  expressed in terms of the dimensionless annihilation-rate parameter $\Zeff$, which is the experimentally measured annihilation rate $\lambda_m$ normalized by the Dirac annihilation rate for a free-electron gas with number density equal to that of the molecular gas \cite{Gribakin10}. Thus, 
\begin{equation}
\Zeff = \frac{\lambda_m}{n_m c \pi r_0^2},
\end{equation}
where $n_m$ is the test-gas number density, $r_0$ is the classical electron radius, and $c$ is the speed of light. 

As mentioned above, for molecules that support a bound state, $\Zeff$ at low incident positron energies is dominated by  VFRs. In the experiments, the annihilation is measured per pulse of positrons. The pulse has a well-defined energy distribution, and so the measured annihilation rate is a convolution of the experimental beam parameters with the energy dependence of $\Zeff$. In practice, the energy width of a VFR is much smaller than the energy width of the beam, and thus each resonance is treated as a delta function positioned at the resonant energy given by Eq.~(\ref{eq:eps_nu}). This also means that the observed line shapes are determined by the energy distribution of the beam.

Assuming that only infrared-active vibrational modes contribute to the VFRs, Gribakin and Lee found the resonant annihilation spectrum as a function of the mean parallel beam energy $\Ebar$ by convolving the beam distribution with delta functions at each of the resonances \cite{Gribakin06}. Generally speaking, although the shape of the $\Zeff$ spectrum is close to such predictions, the magnitude is often quite different. This is a result of the interaction between the single-mode resonances and multimode vibrations, which can lead to both enhancement or suppression of the mode-based VFR \cite{Gribakin10,Gribakin04,Danielson13}. In order to use the theory for the analysis of the experimental results, an adjustable amplitude factor is included for each resonance.

The final formula to compare to the experiments is \footnote{See Eq.~(44) in Ref.~\cite{Gribakin10}. In practice, we also add a small contribution of the direct annihilation, Eq. (29) in Ref.~\cite{Gribakin10}. However, its contribution in negligibly small for most molecules analysed.} 
\begin{equation}\label{eq:Zeff_res}
\Zeff(\Ebar) = 2\pi^2 \delta_{ep} \sum_\nu\frac{\Bn\Gamma_\nu^e}{k_\nu\Gamma_\nu}f(\Eres-\Ebar ),
\end{equation}
where $\delta_{ep}$ is the electron-positron contact density in the bound state [see Eq.~(\ref{eq:cd_kappa_prop}) in Sec.~\ref{sec:theory}], $k_\nu = \sqrt{2\Eres}$, $\Gne$ is the elastic width (related to the infrared activity of the mode $\nu$), $\Gn$ is the total width, $f(\Eres - \Ebar)$ is the beam energy distribution evaluated at the resonance energy, and $\Bn$ is the adjustable amplitude coefficient. The energy $\Ebar $ is the mean parallel energy of the positrons in the beam (see below). In principle, $\Bn$ could be adjusted for each mode. However the data are not sufficient to distinguish individual modes, and so, for the spectra reported here, each group of modes in an energy band is treated identically.

The application of Eq.~(\ref{eq:Zeff_res}) is simplified by the fact that $\Gamma_\nu^e/\Gamma_\nu \approx 1$ for most infrared-active modes except those that are anomalously weak (see, e.g., Refs.~\cite{Gribakin06,Gribakin17}). We thus set this ratio to unity. The $\Zeff $ obtained from Eq.~(\ref{eq:Zeff_res}) by setting $\Bn = 1$ for each mode represents the maximum resonant annihilation rate due to mode-based VFR (i.e., in the absence of nonlinear mode-mixing effects), which we denote $\Zeffm$. Neglecting the difference between $\Gamma_\nu$ and $\Gamma_\nu^e$, this would be the result of the ``pure'' Gribakin-Lee model \cite{Gribakin06}.

The experiments were conducted using a high-energy-resolution positron beam obtained from a buffer-gas trap. This has been described in detail previously \cite{Murphy92,Gribakin10}, and so presented here is only a brief summary of the technique. Low-energy positrons from a $^{22}$Na source and solid-neon moderator are magnetically guided into a three-stage Penning-Malmberg buffer-gas trap (BGT). After accumulation, energy-tunable pulses are formed by carefully ejecting the positrons out of the trapping potential well to produce a minimum spread of particle energies \cite{Gilbert97,Natisin2015,Natisin2016a}. For most experiments, the pulse rate is $\sim $2~Hz, with typically 20\,000 positrons per pulse.

The positron pulses are magnetically guided through a cylindrical electrode that contains a low-pressure test gas ($\sim $10~$\mu$Torr). A high-sensitivity cesium-iodide gamma-ray detector is used to measure the annihilation gammas produced as the positron pulses interact with the gas. The annihilation rate as a function of incident positron energy is obtained by scanning the bias voltage on the gas cell on subsequent pulses. The absolute annihilation rate, $\Zeff$, is obtained using the measured positron number in a pulse, the measured gas pressure, the known path length, and the detector efficiency \cite{Gilbert02,Barnes03}. The uncertainty shown in the plots of $\Zeff$ (see below) is statistical and based solely on finite counting statistics. There is a further systematic uncertainty in the absolute calibration of 10--20\%, mostly set by the pressure calibration.


The beam is characterized by separate parallel and perpendicular energy distributions with respect to the guiding magnetic field. The perpendicular energy distribution is a two-dimensional Maxwellian with standard deviation $\sigma_{\perp}$, which is set by the cooling-gas temperature $T_\perp $ in the BGT \cite{Natisin2014}. The parallel distribution is a Gaussian, with mean energy $\Ebar$ and standard deviation $\sigma_{||}$, and is set by the ejection dynamics of the particles exiting the BGT  (see Refs.~\cite{Natisin2015,Natisin2016} for further details). The parallel distribution is measured using a retarding potential analyzer (RPA). The use of the magnetic moment adiabatic invariant enables measurement of the perpendicular component by measurement of the mean parallel energy with the analyzer at different magnetic fields (see Refs.~\cite{Natisin2015,GhoshBeam} for details). The time of flight of positrons through the RPA is used as a cross check on the measured mean parallel energy of the beam.  

The important consideration for the annihilation measurements is the distribution of the total energy $\varepsilon $, which is a convolution of the perpendicular and parallel energy distributions, described by $f(\varepsilon -\Ebar )$. For the case described here, this is known as an exponentially modified Gaussian (EMG) distribution, with a total width that depends on both $\sigma_{||}$ and $T_\perp$ (see Refs.~\cite{Gribakin10, Natisin2015}).  The peak of the EMG is shifted from the measured mean parallel energy by an amount that also depends on both energy components of the beam, and this needs to be accounted for to obtain the correct measured $\Eb$. Each annihilation-spectrum-data run typically takes  8--12 hours. At a minimum, the RPA is used to measure the mean parallel energy of the beam both before and after each run in order to protect against drift. For experiments before 2013, the drift in the mean could be significant, and so, for those data sets, RPA measurements were taken approximately every 2 hours. Due to improvements in the annihilation cell (see, e.g., Ref.~\cite{Natisin_thesis}), recent experiments have a typical drift of 1~meV or less during a 12-hour data run. 

\begin{figure}
\centering
\includegraphics[scale=0.8]{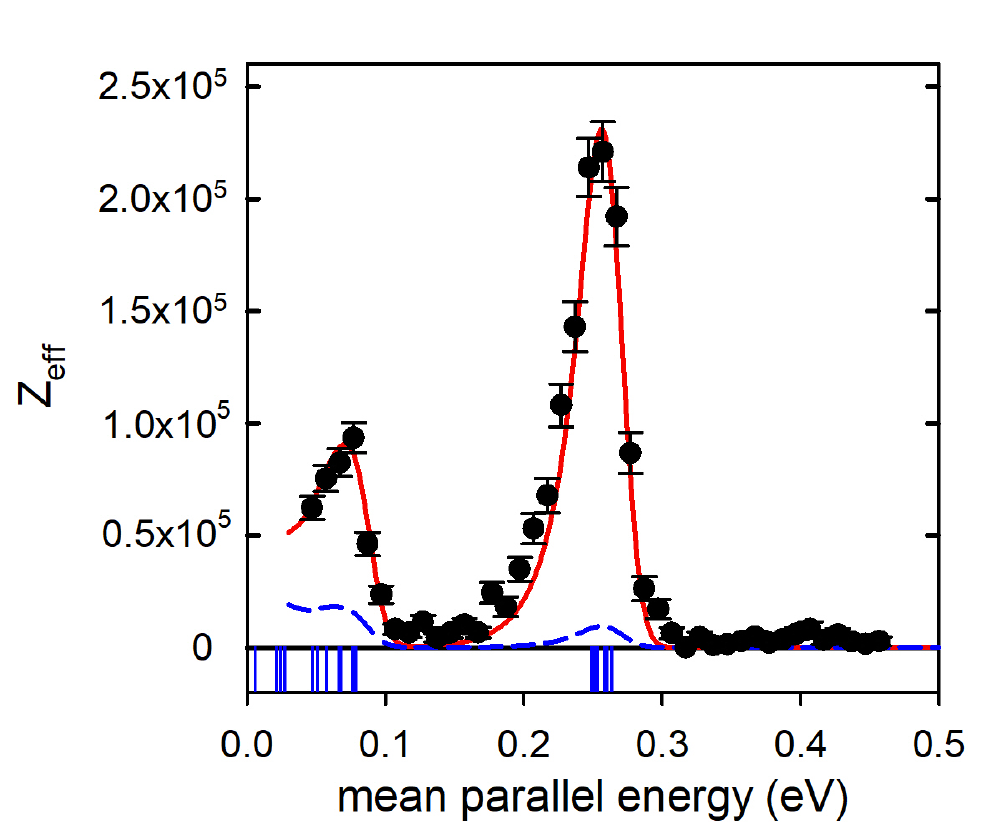}
\caption{\label{fig:data1}
Measured $\Zeff$ for $n$-hexane (circles) with statistical error bars. The solid curve is from the analysis described in the text, resulting in a measured $\Eb = 94 \pm 3$~meV. The dashed curve is $\Zeffm$ obtained using the same $\Eb$. The vertical bars show positions of the infrared-active vibrational modes, downshifted by the sum of $\Eb$ and the peak-to-parallel energy shift for this beam (10~meV).}
\end{figure}

For data taken before 2010,  $\sigma_{||}$  varied from 12 to 15~meV  depending on the experiment, whereas $\sigma_{\perp}$ was typically 25~meV.  For experiments from 2010 to present using the room-temperature BGT, $\sigma_{\perp}$ is typically 20~meV, and $\sigma_{||}$ varies from 9 to 12~meV. These parameters lead to a shift of peak of the total distribution of 10--13~meV. For the cryogenic-beam trap, $\sigma_{\perp}$ is typically 5~meV, and $\sigma_{||}$ varies from 8 to 10~meV. This combination leads to a shift of the peak of 3--4~meV.
In each of Figs.~\ref{fig:data1}--\ref{fig:data3}, which show  measured $\Zeff$ spectra for $n$-hexane, 1-chloropropane, and chloroform (see below), the specific mean-to-peak shift is indicated in the figure caption.

The annihilation-rate data are plotted as $\Zeff$ as a function of the positron mean parallel energy. For the theory fit, each VFR is convolved with the EMG distribution function using the measured beam parameters. The result is a curve that depends solely on the sum of EMG functions, one for each infrared-active vibrational mode, with each shifted to have a peak value at the respective resonance energy [see Eqs.~(\ref{eq:eps_nu}) and (\ref{eq:Zeff_res})]. In this way, $\Eb$ is found by adjusting its value until the resonance locations fit the measured spectrum. For most molecules, the VFR model only qualitatively describes the magnitude of the resonances, and so the amplitudes of the resonances are also adjusted for best fit. Since there are often features beyond just the fundamental modes \cite{Ghosh20}, this procedure is still only approximate. However, in many cases, the main features of the spectrum can be fit well using this procedure, resulting in reliable values of $\Eb$. 

In the present work, this technique is used to obtain values of $\Eb$ for 14 organic molecules that have one or more hydrogen atoms substituted by chlorine atoms. For comparison, updated values for several saturated and unsaturated chain alkanes are also reported.
The data were taken over many years and include results from three different improvements of the experimental apparatus. As described above, each data set had slightly different beam parameters and slightly different procedures used to obtain $\Eb$. In order to
obtain a consistent set of $\Eb$ values and compare them for different molecules, the spectra were corrected for small systematic errors and known differences in the beam parameters and then refit using the procedure described above. 

The data obtained between 2002 and 2006 include the first measurements of the chain-alkane molecules (ethane, propane, butane, and hexane) \cite{Barnes03}, the unsaturated alkanes---acetylene and ethylene \cite{Barnes03}, and methyl chloride \cite{Barnes06}. Molecules in the second data set, taken between 2010 and 2015, include the small molecules---chloroform and carbon tetrachloride \cite{Jones12}, as well as dichloromethane \cite{Natisin_thesis}. Also included from this period are measurements of 1,1-, \textit{trans}-1,2-, and \textit{cis}-1,2-dichloroethylene, and tetrachloroethylene \cite{Natisin_thesis,Gribakin17}. Several of these molecules were later remeasured using the high-resolution cryobeam trap \cite{Natisin17}. The third data set, from 2019, includes 1-chloropropane, 2-chloropropane, and 1,3- and 2,2-dichloropropane. During this period, several of the chain-alkane molecules (including butane and hexane) were remeasured and reanalyzed. These results are included here as minor updates to the previous measurements.

A detailed discussion of all of the data is outside the scope of the current paper. However, we include examples of the measured $\Zeff$ spectra for selected molecules to illustrate the nature of the data and the analysis. Three specific molecules are presented: one is a previously unpublished molecule (1-chloropropane); one is a new measurement of a molecule reported previously (hexane); and one shows the reanalysis correcting a small energy shift (chloroform). The respective energy scales of the new  measurements are believed to be good at the 2--3 meV level, which is the most accurate done to date. As mentioned above, the absolute annihilation-rate values are expected to be good to about 10--20\%.

Shown in Fig.~\ref{fig:data1} is the spectrum of $n$-hexane (C$_6$H$_{14}$), where the measured $\Zeff$ is plotted as a function of the mean parallel energy of the positron beam. As discussed above, the error bars in $\Zeff$ are purely statistical. This molecule was measured before, but the measurements here use slightly improved beam parameters, and the analysis uses a complete set of vibrational modes (i.e., rather than comparing the position of the CH-stretch peak, here at $\sim$0.26~eV, with the average CH-stretch-mode energy \cite{Barnes03}). The solid curve in Fig.~\ref{fig:data1} is the analysis by means of Eq.~(\ref{eq:Zeff_res}) described above, resulting in $\Eb = 94 \pm 3$ meV.  The pure mode-based prediction $\Zeffm$ using the same value for $\Eb$ is shown by the dashed line for comparison. Also shown are the locations of the vibrational modes, downshifted by $\Eb$ and the peak-to-parallel shift of 10 meV. 
Although not perfect, the fit  captures the primary features of the data, including the CH-stretch peak and the sharp, high-energy edge of that peak as well as much of the structure at lower energies. 
Overall, this analysis determines $\Eb$ to within $\pm$3~meV. To obtain a significant change in $\Eb$ would require a large increase in the number of fit parameters, and it would be unlikely to fit both the high- and low-energy portions of the spectrum.

\begin{figure}
\centering
\includegraphics[scale=0.83]{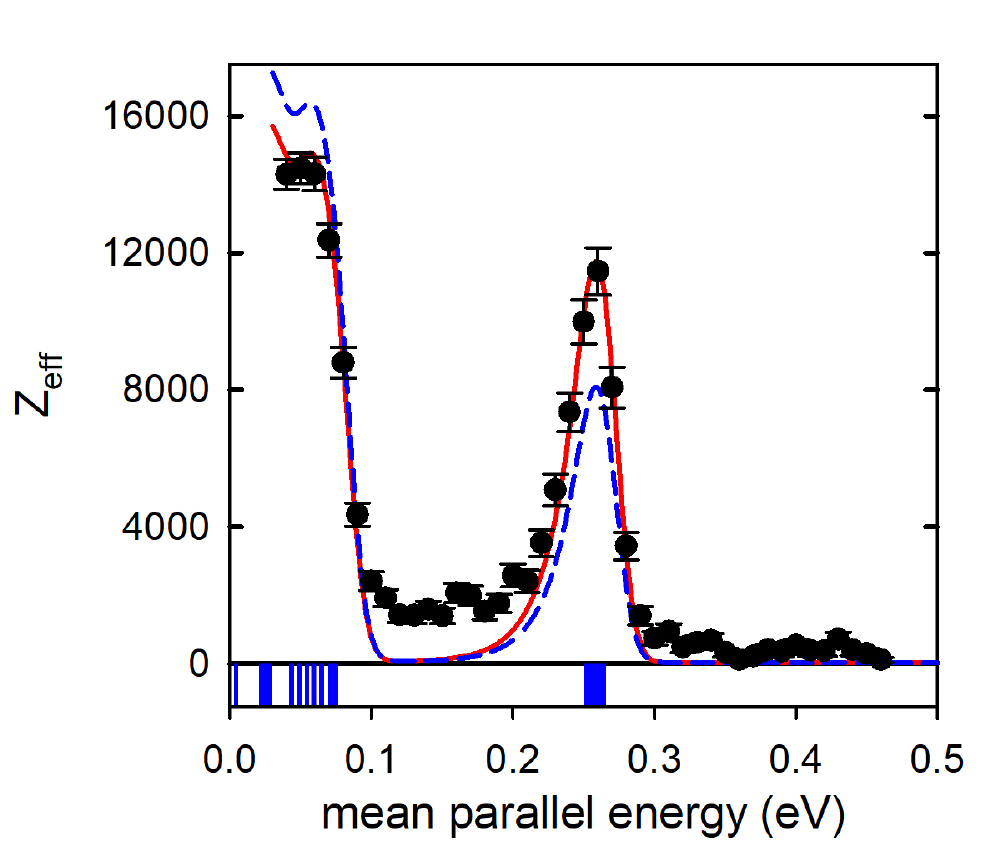}
\caption{\label{fig:data2}
Measured $\Zeff$ for 1-chloropropane (circles) with statistical error bars. The solid and dashed curves and the vertical bars are defined  as in Fig.~\ref{fig:data1}. The analysis gives a measured $\Eb = 98 \pm 4$~meV. As in Fig.~\ref{fig:data1}, the peak-to-parallel shift was 10~meV.
}
\end{figure}

The measurements for 1-chloropropane (or isopropyl chloride, C$_3$H$_7$Cl) are shown in Fig.~\ref{fig:data2}. These data correspond to the same beam parameters as for $n$-hexane. As in Fig.~\ref{fig:data1}, the solid and dashed curves are the complete analysis and $\Zeffm$ predictions, respectively. The analysis yields $\Eb = 98 \pm 4$~meV. For this molecule, although the $\Zeff$ fit describes well the high-energy (CH-stretch) resonance, it misses spectral weight in the range between the high- and low-energy features. The relative error in $\Eb$ is thus slightly larger than for $n$-hexane because of the possible influence of the background on the shape of the resonance in the high-energy region. Nevertheless, the upper edge of the resonance from the lower energy modes provides a strong constraint on any likely variation in $\Eb$.
We note that the raised background is likely due to direct excitation of multimode vibrations \cite{Jones12,Ghosh20}. At the same time, 1-chloropropane is a smaller molecule than $n$-hexane. Its vibrational spectrum is much less dense and there 
is less mixing between single-quantum and multiquantum vibrations, which is why the high- and low-energy peaks are described reasonably well by $\Zeffm$.

\begin{figure}
\centering
\includegraphics[scale=0.8]{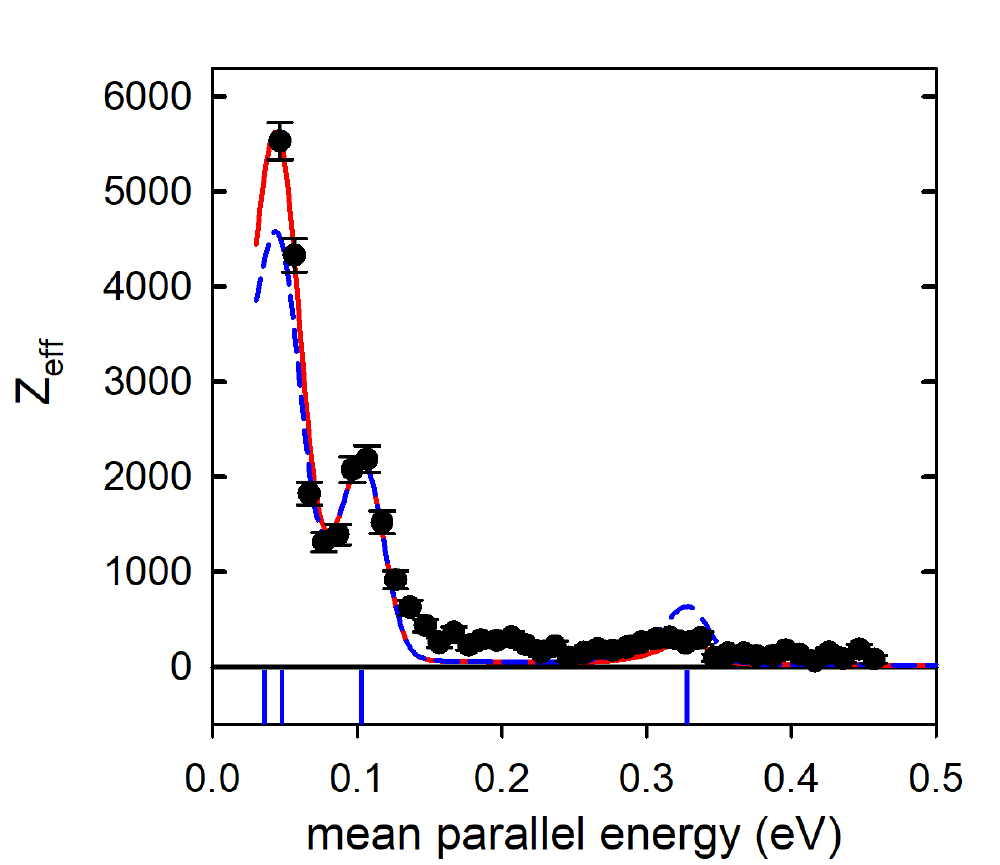}
\caption{\label{fig:data3}
Measured $\Zeff$ for chloroform (circles) with statistical error bars. The data include a small correction to those in Ref.~\cite{Jones13}. The solid and dashed curves and vertical bars are defined as in Fig.~\ref{fig:data1}. The analysis yields $\Eb = 37 \pm 3$~meV. For this data set, the peak-to-parallel shift was 11~meV. 
}
\end{figure}

The spectrum for chloroform (CHCl$_3$) is shown in Fig.~\ref{fig:data3}. The $\Zeff$ for this molecule was originally published in Ref.~\cite{Jones12} and then corrected after remeasurement in Ref.~\cite{Jones13}. The analysis used here includes a small shift in the magnetic field. In this case, the result is $\Eb = 37 \pm 3$~meV. This new value is within the error bars of the previous measurement but with a reduced error bar. For this molecule, the high-energy CH-stretch peak is quite small, but measurable, and so the parameters are well constrained. As also seen in 1-chloropropane, there is extra spectral weight in the intramode region. Such a component has been explained previously as possibly due to a broad spectrum of weak combination and overtone modes \cite{Jones12}. While this hypothesis is still being tested, recent measurements support the idea that combination (i.e., multiquantum) vibrations can contribute to resonant annihilation \cite{Ghosh20}. Here, the amount of this component is relatively small, and so it does not influence significantly the value of $\Eb$.


Values of $\Eb$ obtained from the fits and estimates of the errors are shown in Sec.~\ref{sec:results} (Table~\ref{tab:binding_energies}). All molecules were reanalyzed as described above using the corrected beam parameters to describe the shape of the VFRs. In all cases, the new $\Eb$ are within the previous  $\pm$10~meV uncertainty and in most cases differ by  $<$5~meV. The main effect of the reanalysis is a reduction of the error bars.

The spectra of some of the molecules presented difficulties that led to larger error bars. For some molecules, a systematic shift of the energy scale was discovered \textit{post hoc} (likely due to the electrical-potential variations on the test gas cell). Although the exact values are unknown, the extent of the shifts is bounded by the measured time-of-flight data. Larger error bars are included for these measurements to recognize these additional uncertainties. The spectra of most molecules had several distinct resonances that could be used to constrain the binding energy. However, some only had one or none, and this led to larger error bars for these measurements. To make it clear which data include these difficulties, these measurements are designated with a footnote in Table~\ref{tab:binding_energies}. It should be noted that these specific molecules were not used to fit the model parameters discussed in Sec.~\ref{sec:theory}. However, they are useful results and are included as part of the broader comparisons discussed below.


\subsection{\label{sec:theory}Calculations}

The model-correlation-potential method for calculating positron-molecule binding energies has been described in detail in Ref.~\cite{Swann18}. Here we simply describe the salient features of the method. Atomic units (a.u.) are used throughout this section, except where otherwise stated.

The molecule is treated in the Born-Oppenheimer approximation. Its electronic structure   is described at the static, Hartree-Fock level. Using the standard 6--311++G($d,p$) Gaussian basis, the molecular geometry is optimized, and the doubly occupied electronic orbitals $\varphi_i$ are found. 
The positron-molecule interaction  potential $V$ is taken as the sum of two terms, viz.,
\begin{equation}
V(\mathbf r) = V_\text{st}(\mathbf r) + V_\text{cor}(\mathbf r). \label{eq:potential}
\end{equation}
Here, $V_\text{st}$ is the electrostatic potential of the molecule, given by
\begin{equation}
V_\text{st}(\mathbf r) = \sum_{A=1}^{N_a} \frac{Z_A}{\lvert \mathbf r - \mathbf r_A \rvert}
- 2 \sum_{i=1}^{N/2} \int \frac{\lvert \varphi_i(\mathbf r')\rvert^2}{\lvert \mathbf r - \mathbf r' \rvert} \, d\tau' ,
\end{equation}
where $N_a$ is the number of nuclei in the molecule, $Z_A$ and $\mathbf r_A$ are the charge and position of nucleus $A$, respectively,  and $N$ is the number of electrons. The second term in Eq.~(\ref{eq:potential}), $V_\text{cor}$, represents the electron-positron correlation interaction. Its exact form is provided by many-body theory, but is nonlocal and energy dependent \cite{Dzuba95,Dzuba96,Gribakin04a,Green14}, and would be difficult to compute accurately for larger polyatomic molecules. We therefore use a physically motivated expression for $V_\text{cor}$ \cite{Swann18}, viz.,
\begin{equation}\label{eq:Vcor}
V_\text{cor}(\mathbf r) = - \sum_{A=1}^{N_a} \frac{\alpha_A}{2 \lvert \mathbf r - \mathbf r_A \rvert^4} [1 - \exp(-\lvert\mathbf r - \mathbf r_A \rvert^6/\rho_A^6)] ,
\end{equation}
where $\alpha_A$ are atomic hybrid polarizabilities \cite{Miller90}, and $\rho _A$ are cutoff radii. The hybrid polarizabilities take into account the chemical environment of each atom within the molecule, and the total polarizability of the molecule is $\alpha=\sum_A\alpha_A$. The cutoff factor in square brackets suppresses the  potential at distances close to an atomic nucleus, with $\rho_A$ chosen by comparison with experiment or high-quality calculations. At long range, Eq.~(\ref{eq:Vcor})  describes polarization of the molecule by the positron. The short-range part of $V_\text{cor}$ accounts for other correlation effects, such as virtual positronium formation.

The Schr\"odinger equation for the positron wave function $\psi$ and energy $\varepsilon$, viz.,
$
[-\frac12 \nabla^2 + V(\mathbf r) ]\psi(\mathbf r) = \varepsilon \psi(\mathbf r) ,
$
is solved using a Gaussian basis, which yields a discrete set of positron states. Any negative-energy state describes a positron bound to the molecule; the binding energy $\varepsilon_b$ is related to the energy $\varepsilon$ of the state by $\varepsilon_b=-\varepsilon=\lvert\varepsilon\rvert$. The positive-energy states can, in principle, be used to study positron-molecule scattering and direct annihilation \cite{Swann20}.
In practice, the calculations are carried out using \textsc{gamess} \cite{Schmidt93,Gordon05} with the \textsc{neo} package \cite{Webb02,Adamson08}, which we have modified to include $V_\text{cor}$ \cite{Swann18}.

The values of the atomic hybrid polarizabilities $\alpha_A$ are taken from Ref.~\cite{Miller90}. For carbon atoms, the hybrid polarizabillity depends on how the carbon is bonded to the surrounding atoms, i.e., the hybrid polarizability is different for a carbon atom with four single bonds, with a double bond and two single bonds, and with a triple bond and a single bond. These values are shown in Table \ref{tab:polar_cutoff}.
\begin{table}
\caption{\label{tab:polar_cutoff}Values of the atomic hybrid polarizabilities $\alpha_A$ and cutoff radii $\rho_A$ used in the calculations.}
\begin{ruledtabular}
\begin{tabular}{lD{.}{.}{3}l}
Atom $A$ & \multicolumn{1}{c}{$\alpha_A$ (a.u.)} & \multicolumn{1}{c}{$\rho_A$ (a.u.)} \\
\hline
\\[-0.5em]
\chemfig{C(-[:30,0.75])(-[:150,0.75])(-[:210,0.75])(-[:330,0.75])} & 7.160 & 2.15, 2.25 \\[1.5em]
\chemfig{C(=[:180,0.75])(-[:30,0.75])(-[:330,0.75])} & 9.124 & 2.15, 2.25 \\[1.5em]
\chemfig{C(~[:180,0.75])(-[:0,0.75])} & 8.658 & 2.15, 2.25 \\[0.2em]
\chemfig{H(-[:180,0.75])} & 2.612 & 2.15, 2.25 \\[0.2em]
\chemfig{Cl(-[:180,0.75])} & 15.62 & 2.20, 2.24 
\end{tabular}
\end{ruledtabular}
\end{table}

For simplicity, we take the cutoff radius $\rho_A$ to be the same for C and H atoms. In Ref.~\cite{Swann19}, a value of $\rho_\text{C,H}=2.25$~a.u. was used to calculate positron binding energies for alkane molecules. This value was chosen to reproduce the experimental binding energy for $n$-dodecane (C$_{12}$H$_{26}$) and, overall, it gave values of $\varepsilon_b$ in good agreement with the experimental data for the other alkanes studied (including predicting the emergence of a second bound state for alkanes with 12 or more carbon atoms). However, for small alkanes, the calculated values of $\varepsilon_b$ somewhat underestimated the experimental values. Thus, for propane, the calculation yielded $\varepsilon_b=4.3$~meV \cite{Swann19} vs. the measured value of 10~meV \cite{Young08_large}, while the improved measured value $\varepsilon_b=16\pm 3$~meV (see Sec.~\ref{sec:results}) is even larger. Therefore, in this work we choose the cutoff radius  of $\rho_\text{C,H}=2.15$~a.u., which gives $\varepsilon_b$ in agreement with the improved measurement for propane. An exception is made for the larger molecules $n$-hexane and $n$-hexyl chloride, for which we use the original value of $\rho_\text{C,H}=2.25$~a.u. from Ref.~\cite{Swann19}.
The cutoff radius for a chlorine atom can be chosen to reproduce the measured binding energy for one of the chlorine-containing molecules (using the already fixed value of $\rho_\text{C,H}$). To obtain an insight into the sensitivity of the calculated binding energy to the exact value of $\rho_\text{Cl}$ used, we have performed all calculations for two different values, viz., $\rho_\text{Cl} = 2.20$ and 2.24~a.u. The former is chosen to reproduce the measured binding energies for carbon tetrachloride (CCl$_4$) and $n$-propyl chloride (C$_3$H$_7$Cl), as well as the measured upper bound value of $\varepsilon _b$ in chloroform (CHCl$_3$). The latter fits the lower bound for $\varepsilon _b$ in chloroform. All of these data are summarized in Table \ref{tab:polar_cutoff}.

The Gaussian basis used for solving the Schr\"odinger equation for the positron consists of 12 $s$-type primitives  (i.e., functions of the form $\exp[\zeta^{(A)}_i \lvert \mathbf r - \mathbf r_A \rvert^2]$) centered on each C or Cl nucleus, and eight $s$-type primitives centered on each H nucleus. The exponents $\zeta^{(A)}_i$ of the primitives centered on each nucleus $A$ are chosen according to an even-tempered scheme: $\zeta^\text{(C,Cl)}_i=0.0001 \times 3^{i-1}$ ($i=1$--12) and $\zeta^\text{(H)}_i=0.0081\times 3^{i-1}$ ($i=1$--8).
Since the true wave function of the bound positron behaves at large distances as $\psi(\mathbf r) \sim e^{-\kappa r}$, where $\kappa=\sqrt{2\varepsilon_b}$, our choice of 0.0001 for the smallest exponent of the basis means that bound states with $2\varepsilon_b = \kappa^2 \gtrsim 0.0001$, i.e., $\varepsilon_b \gtrsim 1$~meV, will be well described.
The use of higher-angular-momentum-type Gaussian primitives would improve the description of the bound positron state, but only slightly, since the placement of $s$-type functions on multiple centers effectively generates higher-angular-momentum-type functions \cite{Whitten63,Petke69,Swann18}.
For example, previous calculations for positron binding to hydrogen cyanide showed that, compared to using only $s$-type functions, including $p$- and $d$-type functions in the basis increased the binding energy by only 2--3\% \cite{Swann18}. Such a small change in the binding energy is within the uncertainty due the choice of the values of the cutoff radii.

In addition to the binding energy, we also calculate the electron-positron contact density $\delta_{ep}$ in the bound state. This determines the annihilation rate of the positron-molecule complex,
\begin{equation}
\Gamma = \pi r_0^2 c \delta_{ep} , \label{eq:ann_rate}
\end{equation}
where $r_0$ is the classical electron radius and $c$ is the speed of light \cite{annrate_note,Gribakin10}. The contact density is defined by
\begin{equation}
\delta_{ep} =  \int  \sum_{i=1}^N \delta(\mathbf r - \mathbf r_i) \lvert \Psi(\mathbf r_1  , \dotsc , \mathbf r_N , \mathbf r) \rvert^2 \, d\tau_1  \dotsm d\tau_N \, d\tau , \label{eq:contact_density}
\end{equation}
where $\mathbf r_i$ is the position of the $i$th electron ($i=1,\dotsc,N$), $\mathbf r$ is the position of the positron, and $\Psi$ is the wave function of the $N+1$ particles, normalized as 
\begin{equation}
\int \lvert \Psi(\mathbf r_1  , \dotsc , \mathbf r_N , \mathbf r) \rvert^2  \, d\tau_1  \dotsm d\tau_N \, d\tau=1.
\end{equation}
For atoms and nonpolar molecules with ionization potentials greater than the positronium binding energy ($6.8$~eV), the contact density scales approximately as the square root of the binding  energy, i.e.,
\begin{equation}
\delta_{ep} \simeq \frac{F}{2\pi} \kappa , \label{eq:cd_kappa_prop} 
\end{equation}
where $F$ is a constant ($F\approx 0.66$ for positron-atom bound states \cite{Gribakin01}). For polyatomic molecules, this scaling is confirmed by calculations of the contact-density values for alkanes \cite{Swann19}.

In the present approach, the wave function of the positron-molecule system is of the form
\begin{equation}
\Psi(\mathbf r_1  , \dotsc , \mathbf r_N , \mathbf r) = \Phi(\mathbf r_1  , \dotsc , \mathbf r_N) \psi(\mathbf r) , \label{eq:wf_factoriz}
\end{equation}
where $\Phi(\mathbf r_1  , \dotsc , \mathbf r_N)$ is a Slater determinant of the $N$ electronic spin orbitals, and $\psi$ is the positron wave function. Using Eq.~(\ref{eq:wf_factoriz}) in Eq.~(\ref{eq:contact_density}) gives the following expression for the contact density:
\begin{equation}
\delta_{ep}^{(0)} =  \sum_{i=1}^{N/2} \rho_i , \label{eq:cd_unenh}
\end{equation}
where 
\begin{equation}
\rho_i = 2 \int \lvert \varphi_i(\mathbf r)\rvert^2 \lvert \psi(\mathbf r)\rvert^2 \, d\tau
\end{equation}
is the overlap of the positron density with the electron density in the doubly occupied spatial electronic orbital $\varphi_i$. In Eq.~(\ref{eq:cd_unenh}), the superscript $(0)$ has been added to $\delta_{ep}$ to indicate that this estimate of the contact density should be considered as a zeroth-order approximation; it significantly underestimates the true value. The product form of the wave function (\ref{eq:wf_factoriz}) does not account for strongly attractive short-range correlations that increase the density of the electrons at the positron.
In many-body theory, these correlations are represented by the annihilation-vertex corrections \cite{Gribakin04a,Dunlop06,Green14,Green15}. Their effect can be taken into account by introducing orbital-specific enhancement factors $\gamma_i\geq1$ into the calculation of the contact density \cite{Green15},
\begin{equation}
\delta_{ep} =  \sum_{i=1}^{N/2} \gamma_i \rho_i . \label{eq:cd_enh}
\end{equation}
Many-body-theory calculations of positron-atom annihilation have shown that $\gamma_i$ depends on the ionization energy of the orbital, $I_i$, and can be approximated by the formula \cite{Green15}:
\begin{equation}
\gamma_i =1 + \sqrt{1.31 / I_i} + \left( 0.834 / I_i \right)^{2.15} . \label{eq:enh_fac}
\end{equation}


\section{\label{sec:results}Results}

The calculated and measured binding energies are shown in Table \ref{tab:binding_energies}.
\begin{table*}
\caption{\label{tab:binding_energies}Calculated and measured positron binding energies for the molecules studied. Also shown are the  calculated values of the molecular dipole moment $\mu$ and polarizability $\alpha$ used in the calculations.}
\begin{ruledtabular}
\begin{tabular}{llD{.}{.}{4}cD{.}{.}{2}D{.}{.}{2}c}
&&&& \multicolumn{1}{c}{$\varepsilon_b^{+}$} & \multicolumn{1}{c}{$\varepsilon_b^{-}$} & \multicolumn{1}{c}{$\varepsilon_b^\text{exp}$}  \\
Molecule &  Formula & \multicolumn{1}{c}{$\mu$ (D)}  & $\alpha$ (a.u.) & \multicolumn{1}{c}{(meV)} & \multicolumn{1}{c}{(meV)} & \multicolumn{1}{c}{(meV)} \\
\hline
Methane & CH$_4$ & 0 & 17.61 & \multicolumn{2}{c}{$-3.021$\footnotemark[2]}    \\
Methyl chloride & CH$_3$Cl & 2.31 & 30.62 & 28.91 & 26.00 & \phantom{0}$26 \pm 6$ \\ 
Methylene chloride & CH$_2$Cl$_2$ & 2.01 & 43.62 & 33.90 & 29.66 & \phantom{0}$32 \pm 4$ \\ 
Chloroform & CHCl$_3$ & 1.33 & 56.63 & 40.31 & 34.31 & \phantom{0}$37 \pm 3$ \\ 
Carbon tetrachloride & CCl$_4$ & 0 & 69.64 & 55.34 & 46.68 & \phantom{0}$55 \pm 10$\footnotemark[1] \\[0.5em] 
Acetylene & C$_2$H$_2$ & 0 & 22.54 & \multicolumn{2}{c}{$-0.8408$\footnotemark[2]}  & $\negphantom{\gtrsim}\phantom{\footnotemark[2]}\phantom{00}{\gtrsim}0$\footnotemark[3] \\ 
Chloroacetylene & C$_2$HCl & 0.693 & 35.55 & 14.36 & 12.76       \\
Dichloroacetylene & C$_2$Cl$_2$ & 0 & 48.56 & 26.75 & 23.63       \\[0.5em]
Ethylene & C$_2$H$_4$ & 0 & 28.70 & \multicolumn{2}{c}{4.802} & \phantom{0}$20 \pm 10$\footnotemark[1] \\ 
Vinyl chloride & C$_2$H$_3$Cl & 1.81 & 41.70 & 54.08 & 49.78  \\
\textit{trans}-1,2-Dichloroethylene & C$_2$H$_2$Cl$_2$ & 0 & 54.71 & 28.84 & 24.89 & \phantom{0}$14 \pm 3$ \\ 
Vinylidene chloride & C$_2$H$_2$Cl$_2$ & 1.75 & 54.71 & 79.24 & 71.89 & \phantom{0}$30 \pm 5$ \\ 
\textit{cis}-1,2-Dichloroethylene & C$_2$H$_2$Cl$_2$ & 2.30 & 54.71 & 107.4 & 98.68  & \phantom{0}$66 \pm 10$\footnotemark[1] \\ 
Trichloroethylene & C$_2$HCl$_3$ & 1.10 & 67.72 & 84.04 & 75.04 & \phantom{0}$50 \pm 10$\footnotemark[1] \\ 
Tetrachloroethylene & C$_2$Cl$_4$ & 0 & 80.73 & 103.1 & 91.66 & \phantom{0}$57 \pm 6$ \\[0.5em] 
Ethane & C$_2$H$_6$ & 0 & 29.99 & \multicolumn{2}{c}{$-1.395$\footnotemark[2]}  & $\negphantom{\gtrsim}\phantom{\footnotemark[2]}\phantom{00}{\gtrsim}0$\footnotemark[3]  \\
Ethyl chloride  & C$_2$H$_5$Cl & 2.48 & 43.00 & 72.95 & 67.74   \\
Ethylene dichloride & C$_2$H$_4$Cl$_2$ & 0 & 56.01 & 13.87 & 11.22 \\
Ethylidene chloride & C$_2$H$_4$Cl$_2$ & 2.42 & 56.01 & 89.59 & 82.02  \\[0.5em]
Propane & C$_3$H$_8$ & 0.0857 & 42.38 & \multicolumn{2}{c}{14.33} & \phantom{0}$16 \pm 3$ \\ 
$n$-Propyl chloride & C$_3$H$_7$Cl & 2.54 & 55.38 & 100.7 & 94.71 & \phantom{0}$98 \pm 4$ \\ 
Isopropyl  chloride & C$_3$H$_7$Cl & 2.57 & 55.38 & 130.0 & 122.7  & $113 \pm 5$ \\ 
1,2-Dichloropropane & C$_3$H$_6$Cl$_2$ & 0.675 & 68.39 & 52.65 & 47.38 \\
1,3-Dichloropropane & C$_3$H$_6$Cl$_2$ & 2.23 & 68.39 & 66.80 & 61.69 & \phantom{0}$85 \pm 10$\footnotemark[1] \\ 
1,1-Dichloropropane & C$_3$H$_6$Cl$_2$ & 2.58 & 68.39 & 131.9 & 122.6 \\
2,2-Dichloropropane & C$_3$H$_6$Cl$_2$ & 2.65 & 68.39 & 158.5 & 147.8 & $131 \pm 4$ \\[0.5em] 
$n$-Butane & C$_4$H$_{10}$ & 0 & 54.76 & \multicolumn{2}{c}{49.49} & \phantom{0}$37 \pm 3$ \\ 
$n$-Butyl chloride & C$_4$H$_9$Cl & 2.63 & 67.77 & 125.2 & 118.9     \\
\textit{sec}-Butyl chloride & C$_4$H$_9$Cl & 2.54 & 67.77 & 181.4 & 172.8   \\[0.5em]
Isobutane & C$_4$H$_{10}$ & 0.126 & 54.76 & \multicolumn{2}{c}{51.04} & $41 \pm 3$ \\
Isobutyl chloride & C$_4$H$_9$Cl & 2.47 & 67.77 &  145.3 & 138.2      \\
\textit{tert}-Butyl chloride & C$_4$H$_9$Cl & 2.63 & 67.77 & 195.5 & 186.1  \\[0.5em]
$n$-Hexane & C$_6$H$_{14}$ & 0 & 79.53 & \multicolumn{2}{c}{85.85} & \phantom{0}$94 \pm 3$ \\
$n$-Hexyl chloride & C$_6$H$_{13}$Cl & 2.70 & 92.54 & 133.6 & 127.7 & $170 \pm 15$\footnotemark[1] 
\end{tabular}
\end{ruledtabular}
\footnotetext[1]{These measurements have large error bars due to additional uncertainties in the measurements. See text for details.}
\footnotetext[2]{Negative values indicate that there is no bound state for the positron. The magnitude of the value shown is the lowest energy eigenvalue of the positron Hamiltonian; it is determined by the size of the basis.}
\footnotetext[3]{``${\gtrsim}0$'' indicates that the measured $\Zeff$ displays VFR, which indicates binding, but the binding energy is too small to measure.}
\end{table*}
The values calculated using $\rho_\text{Cl}=2.20$~a.u. and 2.24~a.u. are denoted by $\varepsilon_b^{+}$ and $\varepsilon_b^{-}$, respectively; the smaller cutoff radius gives a larger binding energy and \textit{vice versa}. For pure hydrocarbons, there is only one value. The measured values are denoted by $\varepsilon_b^\text{exp}$.
For each molecule, the table also shows the  permanent dipole moment $\mu$ generated by the static electron distribution \footnote{The dipole moments are from the Hartree-Fock calculations by \textsc{gamess}. They may be slightly larger than the true values, but are sufficient for a qualitative analysis of the results.}, and the polarizability $\alpha=\sum_A \alpha_A$. Below we first discuss the calculated binding energies and then compare theoretical binding energies with experiment.

We begin by noting that the calculated binding energies are relatively insensitive to the choice of $\rho_\text{Cl}$: the difference between $\varepsilon_b^{+}$ and $\varepsilon_b^{-}$ ranges from 1.6 to 11.4~meV, the average difference being 6.5~meV. 
We now discuss the trends in the calculated binding energies group by group.

\begin{figure*}
\centering
\includegraphics{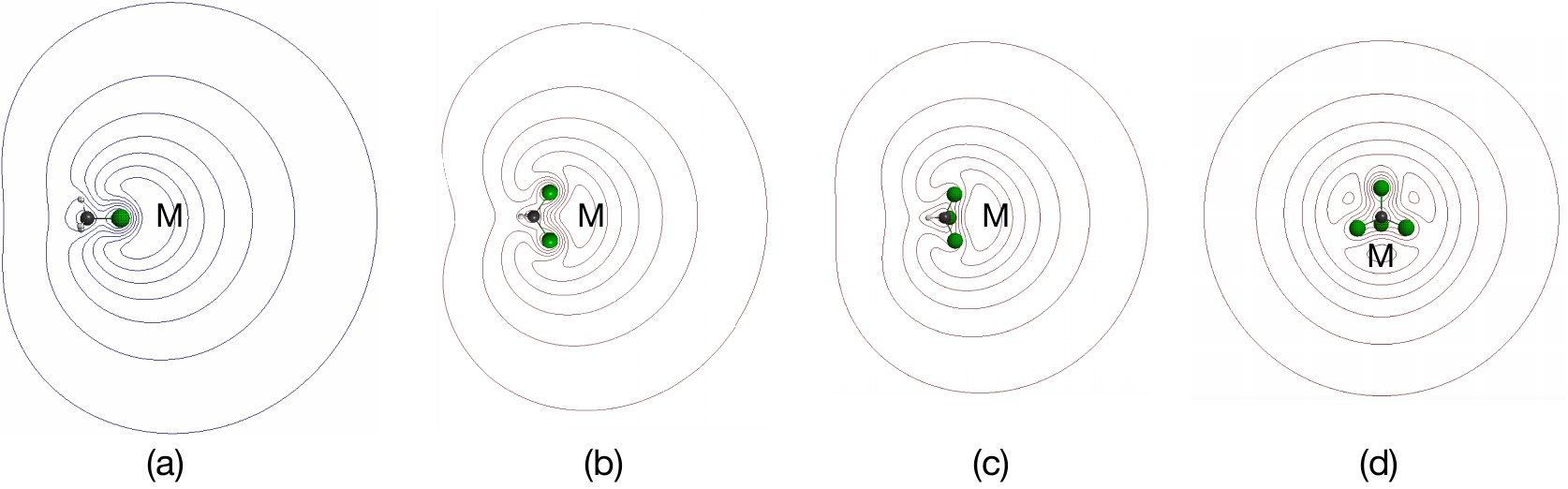}
\caption{\label{fig:chlorinated_methanes}Wave function $\psi$ of the positron bound state for (a) methyl chloride, (b) methylene chloride, (c) chloroform, and (d) carbon tetrachloride, with $\rho_\text{Cl}=2.24$~a.u. The contour with the largest value of the wave function ($\psi=0.014$~a.u.) is marked ``M''. The difference in the value of the wave function between neighboring contours is $\Delta=0.002$~a.u.}
\end{figure*}

Methane is not expected to have a bound state for the positron. For the chlorinated methanes, increasing the number of Cl atoms from one to four steadily increases the binding energy from 26--29~meV to 47--55~meV. This is chiefly due to the increasing polarizability of the molecule, though the binding by methyl chloride is helped a bit by its dipole moment $\mu = 2.31$~D. Figure~\ref{fig:chlorinated_methanes} shows the wave function of the bound positron for the chlorinated methanes, for $\rho_\text{Cl}=2.24$~a.u.
For the polar species,  the positron is most likely to be found outside the negative end of the dipole, i.e., near the Cl atom(s). As the number of Cl atoms increases, the positron becomes more tightly bound and the overall spatial extent of the wave function  decreases. The wave function also becomes more spherically symmetric (since the dipole moment gets smaller).

For acetylene, the calculations with $\rho _\text{C,H}=2.15$~a.u. fail to predict binding, but chlorination leads to binding. The binding energy of dichloroacetylene is almost double that of chloroacetylene. This is due to dichloroacetylene's 40\% larger polarizability, even though it is nonpolar, while chloroacetylene is weakly polar ($\mu=0.693$~D).

Ethylene is predicted to have a bound state with $\varepsilon_b\approx5$~meV. 
For singly chlorinated ethylene, i.e., vinyl chloride, the binding energy is 10 times greater at 50--54~meV. For the doubly chlorinated ethylenes, it becomes clear that the binding energies are determined by more than just the molecular polarizability: the three doubly chlorinated ethylenes all have the same polarizability, yet their binding energies differ greatly, with $\varepsilon_b=25$--29~meV for \textit{trans}-1,2-dichloroethylene,
$\varepsilon_b=72$--79~meV for vinylidene chloride (1,1-dichloroethylene), and $\varepsilon_b=99$--107~meV for \textit{cis}-1,2-dichloroethylene. There are likely two interrelated reasons for this behavior:
  
(1) \textit{trans}-1,2-dichloroethylene is nonpolar, while $\mu=1.75$~D for vinylidene chloride and $\mu=2.30$~D for \textit{cis}-1,2-dichloroethylene; a greater dipole moment leads to greater attraction for the positron at the negative end of the dipole; 

(2) the nonpolar molecule has the two Cl atoms on opposite sides of the molecule, while the two polar molecules have both Cl atoms on the same side. The latter allows the positron to approach both of the highly polarizable Cl atoms at once. 

 We see that changing the geometry of the molecule affects both the electrostatic potential $V_\text{st}$ and the correlation potential $V_\text{cor}$. When the two Cl atoms are on the same side of the molecule, the molecule has  a sizeable dipole moment, which increases the attractive strength of $V_\text{st}$ at the negative (Cl) end of the dipole. The proximity of the two highly polarizable Cl atoms also makes $V_\text{cor}$ strongly attractive here.

\begin{figure*}
\centering
\includegraphics{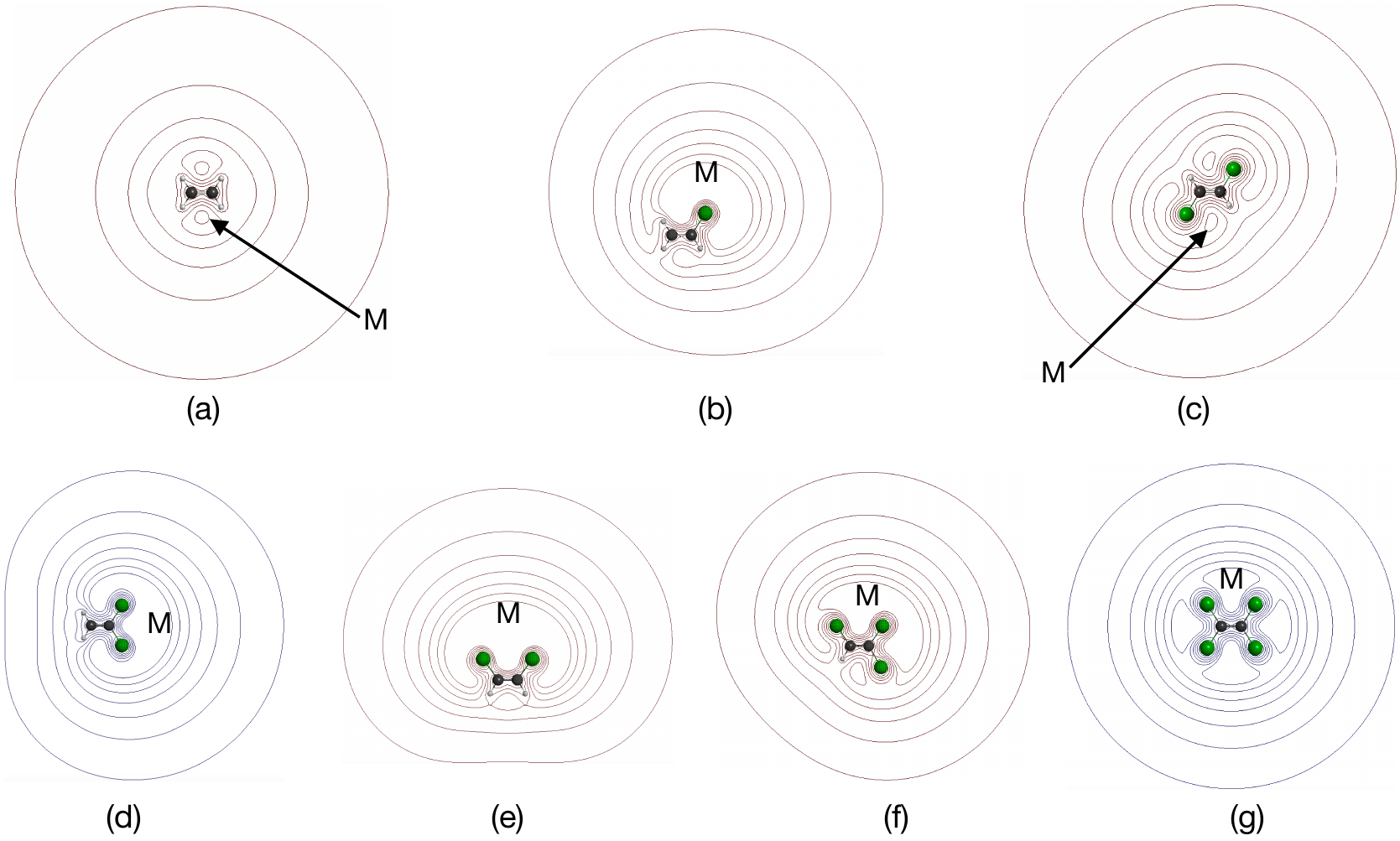}
\caption{\label{fig:chlorinated_ethylenes}Wave function $\psi$ of the positron bound state for (a) ethylene, (b) vinyl chloride, (c) \textit{trans}-1,2-dichloroethylene, (d) vinylidene chloride, (e) \textit{cis}-1,2-dichloroethylene, (f) trichloroethylene, and (g) tetrachloroethylene, with $\rho_\text{Cl}=2.24$~a.u.
The contour for which the value of the wave function is largest ($\psi=0.012$~a.u., except for ethylene, for which $\psi=0.0103$~a.u.) is marked ``M''. The difference in the value of the wave function between neighboring contours is $\Delta=0.0017$~a.u.}
\end{figure*}

The binding energy for trichloroethylene, 75--84~meV, is similar to the binding energy for vinylidene chloride; this is a consequence of trichloroethylene having a larger polarizability but smaller dipole moment than vinylidene chloride.
Similarly, the binding energy for tetrachloroethylene is close to that for \textit{cis}-1,2-dichloroethylene.
The wave functions of the positron bound states for ethylene and the chlorinated ethylenes, with $\rho_\text{Cl}=2.24$~a.u., are shown in Fig.~\ref{fig:chlorinated_ethylenes}.
Again, for molecules with a larger positron binding energy, the wave function of the bound state has a smaller overall spatial extent, and for polar molecules, the positron is most likely to be found near the negative end of the dipole. 
For the nonpolar  species, the maxima of the positron wave function are found to lie above and below the plane of the molecule. 
This is illustrated in Fig.~\ref{fig:ethylene_cis-1,2-DCE}(a) which shows  isosurfaces of the  wave function of the bound positron for ethylene; the value of the wave function on the cyan surfaces  is $\psi=0.012$~a.u., which is close to the maximum value of $\psi=0.0125$~a.u.
\begin{figure}
\centering
\includegraphics[width=\columnwidth]{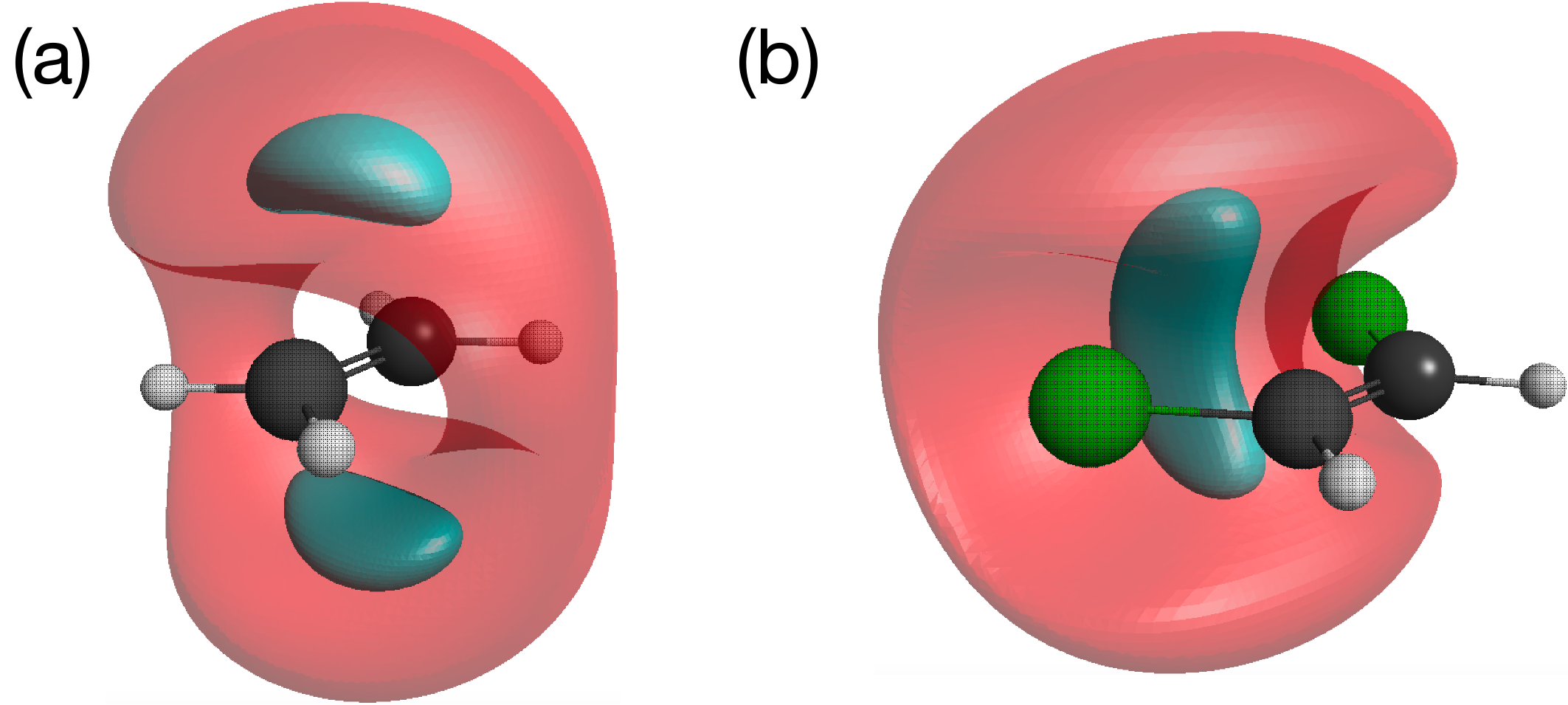} 
\caption{\label{fig:ethylene_cis-1,2-DCE}Wave function $\psi$ of the bound positron  for (a)  ethylene and (b) \textit{cis}-1,2-dichloroethylene, with $\rho_\text{Cl}=2.24$~a.u. 
The values of the wave function on the surfaces shown are as follows. 
Ethylene: red, $\psi=0.010$~a.u.; cyan, $\psi=0.012$~a.u.
\textit{cis}-1,2-Dichloroethylene: red, $\psi=0.016$~a.u.; cyan, $\psi=0.022$~a.u. 
}
\end{figure}
On the other hand, for the polar species, the maxima lie near the negative end of the dipole. Figure \ref{fig:ethylene_cis-1,2-DCE}(b) shows this for \textit{cis}-1,2-dichloroethylene; the value of the wave function on the cyan surface is $\psi=0.022$~a.u., which is close to the maximum value of $\psi=0.0235$~a.u. (which occurs in the plane of the molecule).

For ethane, the calculations with $\rho _\text{C,H}=2.15$~a.u. do not predict binding,
in spite of its polarizability being slightly larger than that of ethylene \footnote{For ethane, the experiments have measured VFRs associated with the vibrational modes, which indicates that it does bind a positron, but the binding energy is small, $\varepsilon _b < 2$ meV \cite{Gilbert02}.}. This may be because ethylene is a planar molecule, while ethane is not: for ethylene, the positron can closely approach the polarizable C atoms from above and below the plane of the molecule, but for ethane, the C atoms are shielded from the positron by the surrounding protons;
moreover, the hybrid polarizability of each C atom in ethane is 22\% smaller than in ethylene (see Table \ref{tab:polar_cutoff}). Replacing one H by Cl leads to binding with $\varepsilon _b\approx 70$~meV. The nonpolar ethylene dichloride (1,2-dichlorethane) and polar ethylidene chloride (1,1-dichloroethane)  have the same polarizability but very different binding energies, similar to the chlorinated ethylenes.

\begin{figure}
\centering
\includegraphics[width=\columnwidth]{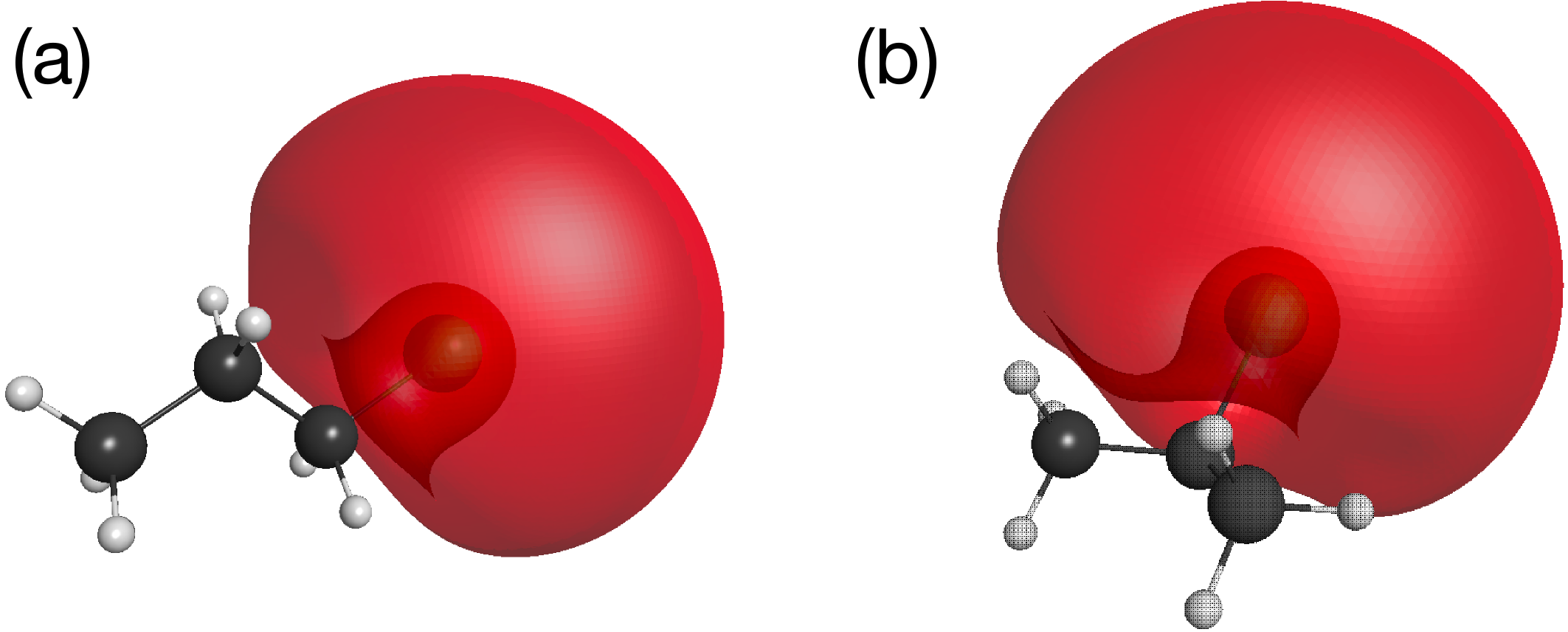} 
\caption{\label{fig:n-vs-iso-propyl-chloride}Wave function $\psi$ of the bound positron  for (a)  $n$-propyl chloride and (b) isopropyl chloride, with $\rho_\text{Cl}=2.24$~a.u. 
The value of the wave function on the surfaces shown is $\psi=0.015$~a.u. }
\end{figure}

Propane is the smallest alkane for which calculations using $\rho _\text{C,H}=2.15$~a.u. predict binding. $n$-Propyl chloride (1-chloropropane) and isopropyl chloride (2-chloropropane) have the same polarizability and almost the same dipole moments ($\mu=2.54$~D vs. $\mu=2.57$~D), yet the binding energy for isopropyl chloride is 30\% larger than that for $n$-propyl chloride. This is a short-range effect due to the difference in the local environment near the chlorine atom, which is the most attractive part of the molecule for the positron. For $n$-propyl chloride, the Cl atom is at the end of the carbon chain; the positron can easily approach the nearest C atom, but its interaction with the other C atoms is reduced. For isopropyl chloride, the location of the Cl atom is more central and is closer to all three C atoms, making it easier for the positrons to interact with them and leading to stronger binding. This is illustrated by Fig.~\ref{fig:n-vs-iso-propyl-chloride}.

\begin{figure*}
\centering
\includegraphics{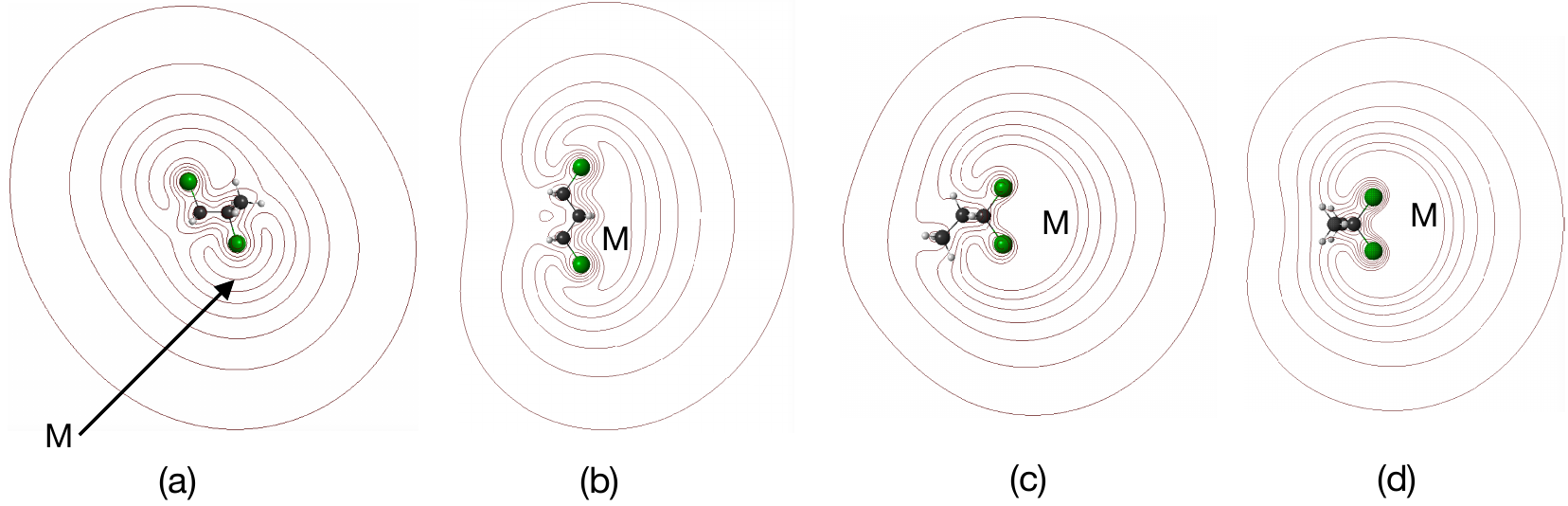}
\caption{\label{fig:chlorinated_propanes}Wave function $\psi$ of the positron bound state for (a) 1,2-dichloropropane, (b) 1,3-dichloropropane, (c) 1,1-dichloropropane, and (d) 2,2-dichloropropane, with $\rho_\text{Cl}=2.24$~a.u. The contour for which the value of the wave function is largest ($\psi=0.014$~a.u.) is marked ``M''. The difference in the value of the wave function between neighboring contours is $\Delta=0.02$~a.u.}
\end{figure*}

For the doubly chlorinated propanes,  all four molecules have the same polarizability, but again there is significant variation in the binding energies. The binding energies for
1,2-dichloropropane (47--53~meV) and 1,3-dichloropropane (62--67~meV) are quite close,
with the difference related to the greater dipole moment of the latter. Putting both Cl atoms on the same carbon atoms increases the binding energy by a factor of 2 or more, to
123--132~meV for 1,1-dichloropropane and 148--159~meV for 2,2-dichloropropane. In the latter case the positron remains in relative proximity to all the C atoms in the molecule, which increases the binding energy. This can be seen from the positron wave functions in Fig.~\ref{fig:chlorinated_propanes}.

Curiously, isopropyl chloride and 1,1-dichloropropane have almost the same binding energy, in spite of the polarizability of the latter being 23\% larger, while their dipole moments are very close (2.57~D vs. 2.58~D). This can be attributed to the difference in the localisation of the positron relative to the molecule. With the positron most likely found off the negative end of the dipole, in 1,1-dichloropropane, it cannot easily embrace the two distant C atoms [see Fig.~\ref{fig:chlorinated_propanes}(c)]; conversely, for isopropyl chloride, the positron is quite close to all three C atoms [see Fig.~\ref{fig:n-vs-iso-propyl-chloride}(b)].

$n$-Butane and isobutane have the same polarizability and almost the same binding energy, $\varepsilon_b\approx50$~meV. While $n$-butane is nonpolar, isobutane is only weakly polar ($\mu=0.126$~D), and the positron wave function surrounds the entire molecule in both cases.
This is similar to Figs.~\ref{fig:chlorinated_methanes}(d), \ref{fig:chlorinated_ethylenes}(a), \ref{fig:chlorinated_ethylenes}(c),  and \ref{fig:chlorinated_ethylenes}(g), which show the positron wave function for other nonpolar species. The typical extent of the wave function in $n$-butane and isobutane is $r \sim 1/\sqrt{2\varepsilon_b} \approx 16$~a.u. The positron is thus found relatively far from the molecule, where it does not probe the molecular geometry, and the positron-molecule interaction is described by its asymptotic form, $V(\mathbf r) \approx V_\text{cor}(\mathbf r) \simeq -\alpha / 2r^4$, where $r$ is the distance from the molecule to the positron [cf.~Eq.~(\ref{eq:Vcor})]. Similarly, previous calculations for $n$-pentane, isopentane, and neopentane similarly showed that their binding energies  differed by only 4~meV \cite{Swann19}.
{
Thus we see that molecular geometry (i.e., structural isomerism) is very important in determining the binding energy for polar molecules where the positron is localized near the negative end of the dipole. In contrast, for nonpolar or weakly polar molecules, its effect is much smaller, provided the binding energies are not too large and the extent of the positron wave function is large compared with the size of the molecule (see also Ref.~\cite{Swann20a}).}

\begin{figure}
\centering
\includegraphics[width=\columnwidth]{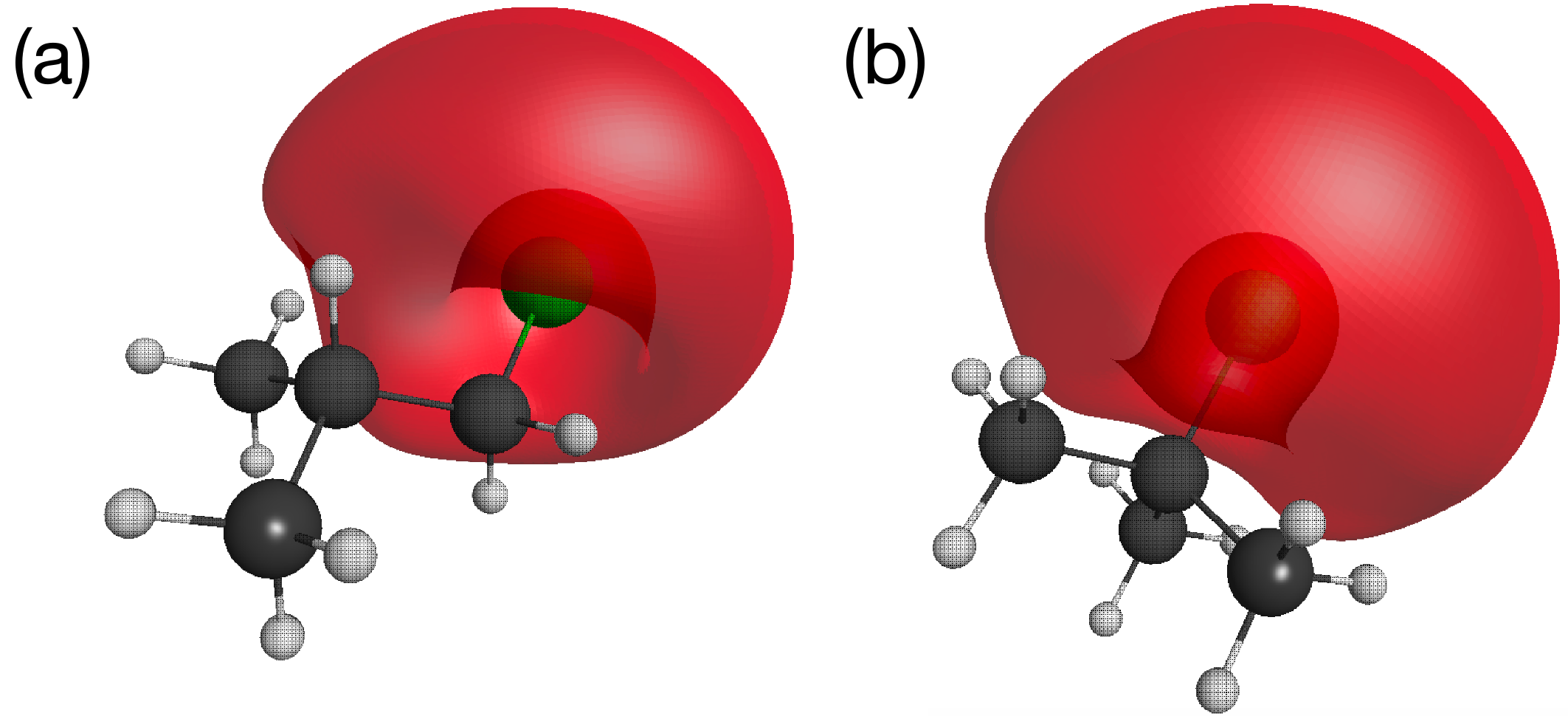} 
\caption{\label{fig:iso-vs-tert-butyl-chloride}Wave function $\psi$ of the bound positron  for (a)  isobutyl chloride and (b) \textit{tert}-butyl chloride, with $\rho_\text{Cl}=2.24$~a.u. 
The value of the wave function on the surfaces shown is $\psi=0.018$~a.u. }
\end{figure}

Replacing a single hydrogen atom by chlorine in $n$-butane or isobutane increases the binding energy by a factor 2.5--4. The singly chlorinated butanes have the same polarizability and similar dipole moments ($\mu=2.47$--2.63~D), yet their binding energies vary greatly. For $n$-butyl chloride where the Cl atom is on the terminal carbon, we find 119--125~meV, while for \textit{tert}-butyl chloride [(CH$_3$)$_3$CCl], we have 186--196~meV, as the positron is close to all four carbon atoms. In isobutyl chloride, the Cl atom is out on the limb, attached to one of the side carbon atoms of isobutane, which reduces the binding energy to 138--145~meV. The difference in the localization of the positron in these two cases can be seen in Fig.~\ref{fig:iso-vs-tert-butyl-chloride}. Finally, \textit{sec}-butyl chloride (173--181~meV) has a greater binding than $n$-butyl chloride (similar to the isopropyl chloride vs. $n$-propyl chloride pair), but not as large as \textit{tert}-butyl chloride.

For nonpolar hexane, the calculation using $\rho _\text{C,H}=2.25$~a.u. gives $\varepsilon_b=86$~meV, while $n$-hexyl chloride (1-chlorohexane, $\mu =2.870$~D) has a binding energy in the range 128--134~meV. This increase is the smallest observed for all 1-chloroalkanes. Here the positron is localized outside the Cl end of the molecule. Viewing $n$-hexyl chloride as part of the sequence of $n$-propyl chloride (95--101~meV) and $n$-butyl chloride (119--125~meV), we see that adding extra carbons at the opposite end of the molecule leads to a slow growth of the binding energy.

\begin{figure}
\includegraphics[width=0.9\columnwidth]{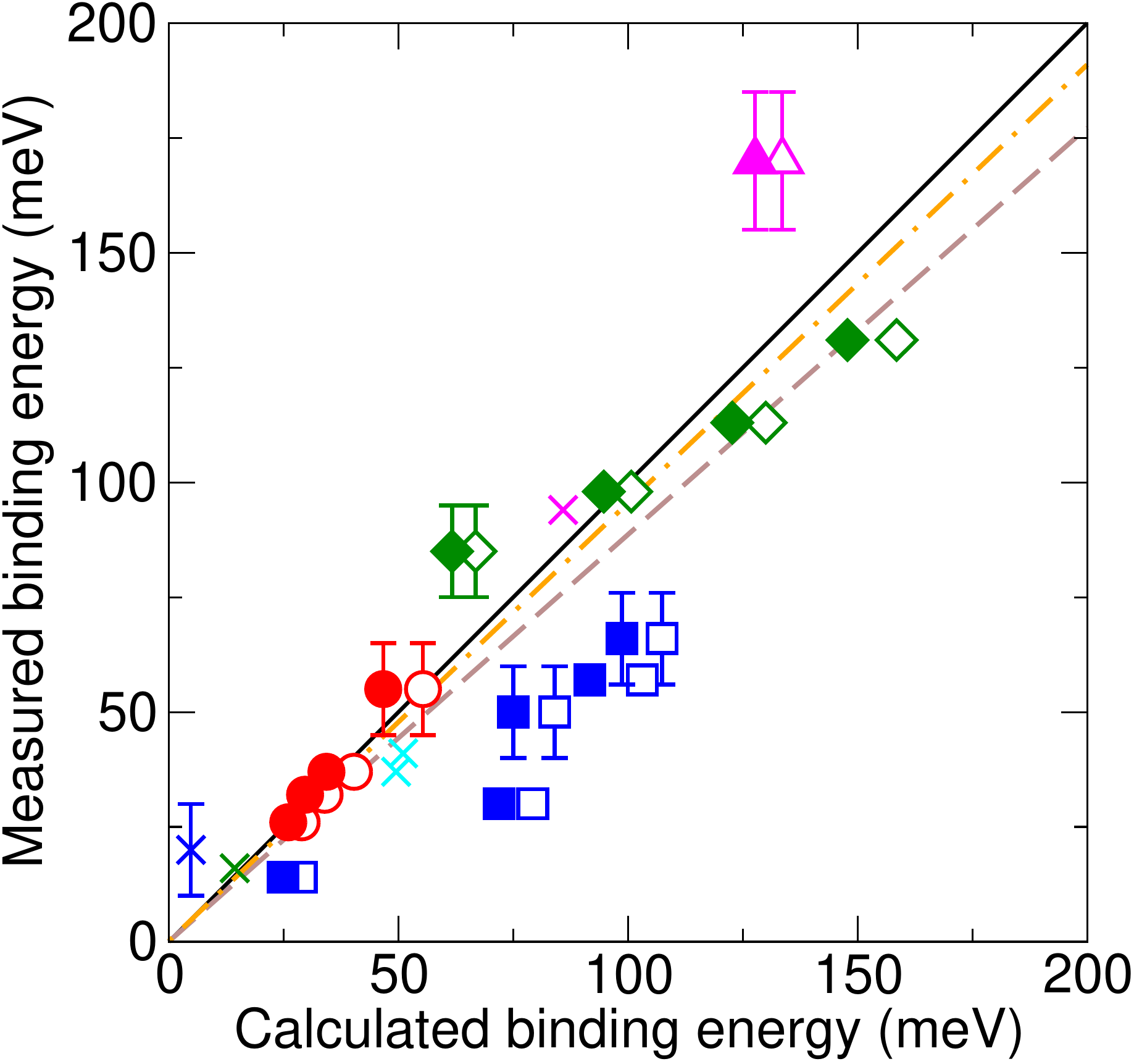}
\caption{\label{fig:exp_vs_theory}Comparison of the measured and calculated binding energies. Crosses are the hydrocarbons, as follows: blue, ethylene; green, propane; cyan, $n$-butane and isobutane; magenta, $n$-hexane. 
Open and filled symbols are the chlorinated hydrocarbons, with $\rho_\text{Cl}=2.20$ and 2.24~a.u., respectively, as follows: red circles, chlorinated methanes; blue squares, chlorinated ethylenes; green diamonds, chlorinated propanes; magenta triangles, $n$-hexyl chloride. Solid black line, identity line; dashed brown line, least-squares fit of measured values to calculated values $\varepsilon_b^{+}$ with $\rho_\text{Cl}=2.20$~a.u.; dot-dashed orange line, least-squares fit of measured values to calculated values $\varepsilon_b^{-}$ with $\rho_\text{Cl}=2.24$~a.u.}
\end{figure}

Figure \ref{fig:exp_vs_theory} shows a comparison of the measured binding energies $\varepsilon_b^\text{exp}$ with the calculated binding energies $\varepsilon_b^{\pm}$.
We see that there is  strong positive correlation between the measured and calculated binding energies. Indeed, the correlation coefficient for the $\varepsilon_b^\text{exp}$ and $\varepsilon_b^{+}$ is 0.863, and for the $\varepsilon_b^\text{exp}$ and $\varepsilon_b^{-}$ it is 0.878. Thus, the calculations with $\rho_\text{Cl}=2.24$~a.u. give the better overall description of the experimental data. Linear regression between the measured and calculated binding energies gives $\varepsilon_b^\text{exp}\approx A_\pm + B_\pm \varepsilon_b^{\pm}$, where $A_+=0.293$, $B_+=0.881$ and $A_-=0.133$, $B_-=0.929$; these lines are also shown in Fig.~\ref{fig:exp_vs_theory}.
Looking at molecular families, the best agreement between theory and experiment is observed for chloromethanes, followed by chloropropanes. In chlorinated ethylenes, there appears to be a systematic difference between the calculated and measured binding energies, though the ordering of calculated and measured binding energies is the same.
We now discuss the possible reasons for the discrepancies observed.

The calculated binding energy for ethylene is four times smaller than the measured value. Conversely, for the chlorinated ethylenes, the calculated $\varepsilon_b^{-}$ are 1.5 to 2.4 times greater than $\varepsilon_b^\text{exp}$. {
The latter is true for both polar and nonpolar chloroethylenes. This indicates that the discrepancy between theory and experiment is likely not due to the use of the Hartree-Fock approximation, which tends to overestimate the molecular dipole moments, for the calculation of $V_\text{st}({\bf r})$.}

A possible deficiency of the present approach is a neglect of anisotropy of the molecular polarizability. At large distances $r$ from the molecule, the chosen form of the positron-molecule correlation potential, Eq.~(\ref{eq:Vcor}), is spherically symmetric, i.e., $ V_\text{cor}(\mathbf r)\simeq -\alpha/2r^4$. However, the true asymptotic form of the positron-molecule interaction is $V_\text{cor}(\mathbf r) \simeq -\sum_{i,j} \alpha_{ij} x_i x_j/2r^6$, where the $x_i$ ($i=1$, 2, 3) are the Cartesian coordinates $x$, $y$, and $z$ of the positron as measured from the molecule, and the $\alpha_{ij}$ are the components of the dipole polarizability tensor \cite{polarizability_note}.

For planar molecules like ethylene and its chloro-substitutes, there can be a significant difference between the components of the polarizability tensor parallel and perpendicular to the plane of the molecule. For example, experiment and calculations of the polarizability tensor for ethylene \cite{Hills75,Hinchliffe97} show that the component perpendicular to the molecular plane is $\alpha_{yy}\approx 0.8\alpha$, while the in-plane component perpendicular to the C$=\joinrel=$C bond is $\alpha_{xx}\approx 0.9\alpha$, and that of the in-plane component parallel to the C$=\joinrel=$C bond is $\alpha _{zz}\approx 1.3\alpha $. Considering the long-range asymmetry of $V_\text{cor}(\mathbf r)$ as a perturbation, and given that the positron wave function for ethylene is spherically symmetric at long range [cf. Fig.~\ref{fig:chlorinated_ethylenes}(a)], such perturbation will result in a zero first-order correction. The second-order perturbative correction to the ground-state energy will be negative, leading to stronger binding for ethylene. For chlorinated ethylenes with positron asymmetric wave functions [e.g., Fig.~\ref{fig:chlorinated_ethylenes}(b), (d), (e) and (f)], the long-range asymmetry of the correlation potential will result in larger first-order corrections.

Qualitatively, in the present calculations, the positron experiences too much attraction above and below the molecular plane, and too little attraction in the molecular plane. At distances comparable to the size of the molecule, for ethylene, the positron is most likely to be found above or below the molecular plane [Fig.~\ref{fig:ethylene_cis-1,2-DCE}(a)]. For \textit{cis}-1,2-dichloroethylene, which is polar, Fig.~\ref{fig:ethylene_cis-1,2-DCE}(b) shows that even though the positron is most likely to be found in the plane of the molecule (between the Cl atoms), the wave function still ``wraps'' around the molecule from above and below the molecular plane.
Thus, the anisotropy of the correlation potential can lead to significant corrections. While it should be possible to take it into account by modifying the form of Eq.~(\ref{eq:Vcor}) (see Sec.~\ref{sec:conc}), this is beyond the scope of the present work.

Another clear outlier is $n$-hexyl chloride. Despite the choice $\rho_\text{C,H}=2.25$~a.u. giving close agreement between the measured and calculated binding energies for $n$-hexane (94~meV vs. 86~meV), the calculated binding energies for $n$-hexyl chloride are 24\% and 27\% (for $\rho_\text{Cl}=2.20$ and 2.24~a.u., respectively) smaller than the measured value. This is in contrast to the good agreement between $\varepsilon _b^\text{exp} =98$~meV and the calculated binding energies of 101 and 95~meV for $n$-propyl chloride.
{
We should note that the CH-stretch peak in the measured $\Zeff$ spectrum for $n$-hexyl chloride is much broader than that in hexane \cite{Young07}, which adds to the uncertainty in the experimental binding energy. Thus, overall, the reason for the discrepancy between theory and experiment for $n$-hexyl chloride remains unclear.}

We now turn to calculations of the electron-positron contact density in the bound state. Since the choice $\rho_\text{Cl}=2.24$~a.u. gives better overall agreement between theory and experiment for the binding energies, we use the corresponding positron wave functions in the calculations of the contact densities.
Table \ref{tab:contact_densities} shows the results.
\begin{table}
\caption{\label{tab:contact_densities}Electron-positron contact densities for $\rho_\text{Cl}=2.24$~a.u. Numbers in brackets indicate powers of 10.}
\begin{ruledtabular}
\begin{tabular}{llccc}
Molecule &  Formula & $\delta_{ep}^{(0)}$ (a.u.) & $\delta_{ep}$ (a.u.) & $\gamma$  \\
\hline
Methyl chloride & CH$_3$Cl & $8.298[-4]$ & $4.190[-3]$ & 5.0   \\
Methylene chloride & CH$_2$Cl$_2$ & $1.075[-3]$ & $5.214[-3]$ & 4.9  \\
Chloroform & CHCl$_3$ & $1.425[-3]$ & $6.722[-3]$ & 4.7     \\
Carbon tetrachloride & CCl$_4$ & $1.974[-3]$ & $9.161[-3]$ & 4.6      \\[0.5em]
%
Chloroacetylene & C$_2$HCl & $8.608[-4]$ & $4.357[-3]$ & 5.1     \\
Dichloroacetylene & C$_2$Cl$_2$ & $1.255[-3]$ & $6.167[-3]$ & 4.9     \\[0.5em]
Ethylene & C$_2$H$_4$ & $6.118[-4]$ & $3.069[-3]$ & 5.0  \\
Vinyl chloride & C$_2$H$_3$Cl & $1.733[-3]$ & $8.602[-3]$ & 5.0  \\
\textit{trans}-1,2-Dichloroethylene & C$_2$H$_2$Cl$_2$ & $1.413[-3]$ & $6.781[-3]$ & 4.8  \\
Vinylidene chloride & C$_2$H$_2$Cl$_2$ & $2.305[-3]$ & $1.112[-2]$ & 4.8 \\
\textit{cis}-1,2-Dichloroethylene & C$_2$H$_2$Cl$_2$  & $2.642[-3]$ & $1.289[-2]$ & 4.9   \\
Trichloroethylene & C$_2$HCl$_3$ & $2.568[-3]$ & $1.225[-2]$ & 4.8  \\
Tetrachloroethylene & C$_2$Cl$_4$ & $3.099[-3]$  & $1.448[-2]$ & 4.7  \\[0.5em]
%
Ethyl chloride  & C$_2$H$_5$Cl & $1.773[-3]$ & $8.876[-3]$ & 5.0  \\
Ethylidene chloride & C$_2$H$_4$Cl$_2$ & $2.185[-3]$ & $1.060[-2]$ & 4.9  \\
Ethylene dichloride & C$_2$H$_4$Cl$_2$ & $8.619[-4]$ & $4.108[-3]$ & 4.8  \\[0.5em]
Propane & C$_3$H$_8$ & $1.213[-3]$ & $5.583[-3]$ & 4.6   \\
$n$-Propyl chloride & C$_3$H$_7$Cl & $2.308[-3]$ & $1.145[-2]$ & 5.0     \\
Isopropyl  chloride & C$_3$H$_7$Cl & $2.909[-3]$ & $1.445[-2]$ & 5.0    \\
1,1-Dichloropropane & C$_3$H$_6$Cl$_2$ & $2.949[-3]$ & $1.424[-2]$ & 4.8  \\
1,2-Dichloropropane & C$_3$H$_6$Cl$_2$ & $2.089[-3]$ & $9.882[-3]$ & 4.7  \\
1,3-Dichloropropane & C$_3$H$_6$Cl$_2$ & $1.898[-3]$ & $9.121[-3]$ & 4.8  \\
2,2-Dichloropropane & C$_3$H$_6$Cl$_2$ & $3.442[-3]$ & $1.667[-2]$ & 4.8 \\[0.5em]
$n$-Butane & C$_4$H$_{10}$ & $2.553[-3]$ & $1.178[-2]$ & 4.6      \\
$n$-Butyl chloride & C$_4$H$_9$Cl & $2.777[-3]$ & $1.362[-2]$ & 4.9     \\
\textit{sec}-Butyl chloride & C$_4$H$_9$Cl & $3.990[-3]$ & $1.960[-2]$ & 4.9  \\[0.5em]
Isobutane & C$_4$H$_{10}$ & $2.681[-3]$ & $1.233[-2]$ & 4.6   \\
Isobutyl chloride & C$_4$H$_9$Cl & $3.401[-3]$ & $1.657[-2]$ & 4.9     \\
\textit{tert}-Butyl chloride & C$_4$H$_9$Cl & $4.170[-3]$ & $2.060[-2]$ & 4.9   \\[0.5em]
$n$-Hexane & C$_6$H$_{14}$ & $3.311[-3]$ & $1.537[-2]$ & 4.6   \\
$n$-Hexyl chloride & C$_6$H$_{13}$Cl & $2.803[-3]$ & $1.373[-2]$ & 4.9 
\end{tabular}
\end{ruledtabular}
\end{table}
We show both the unenhanced contact densities $\delta_{ep}^{(0)}$, given by Eq.~(\ref{eq:cd_unenh}), and the enhanced contact densities $\delta_{ep}$, given by Eqs.~(\ref{eq:cd_enh}) and (\ref{eq:enh_fac}).
The table also shows the overall enhancement factor $\gamma$ for each molecule, defined as $\gamma=\delta_{ep} / \delta_{ep}^{(0)}$.
The value of $\gamma$ is fairly constant across all of the molecules considered, ranging from 4.6 to 5.1. The enhanced contact densities range from $3.1\times10^{-3}$~a.u. for ethylene, to $2.1\times10^{-2}$~a.u. for \textit{tert}-butyl chloride. 
Therefore, the lifetime $1/\Gamma$ of the bound positron-molecule complex ranges from 0.96~ns for \textit{tert}-butyl chloride, to 6.5~ns for ethylene.
This is consistent with the fact that ethylene and \textit{tert}-butyl chloride have the smallest and largest binding energies, respectively, of all of the molecules studied: in general, the more strongly bound the positron is, the more overlap its density will have with the electron density, leading to more rapid annihilation.

\begin{figure}
\includegraphics[width=\columnwidth]{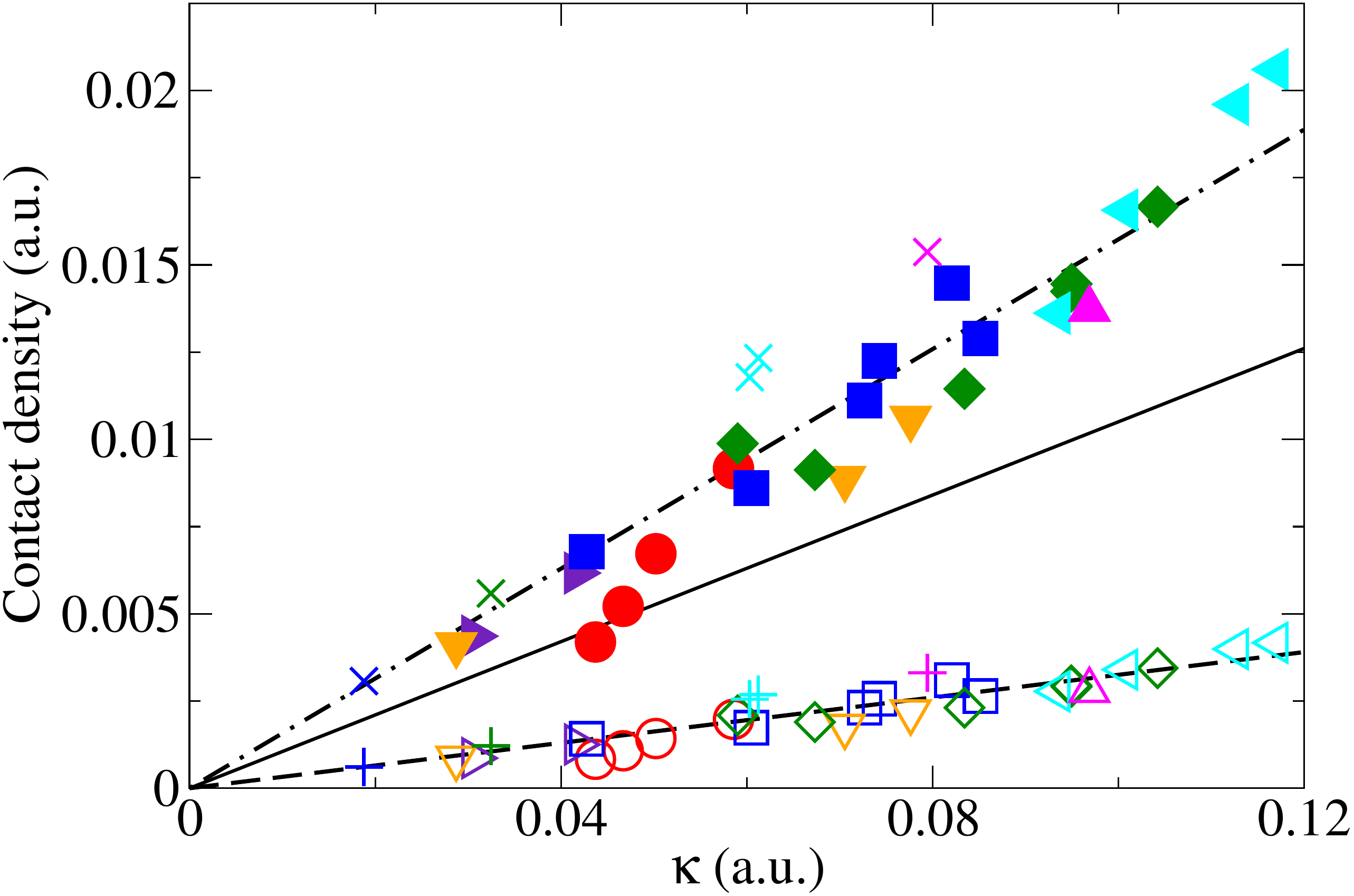}
\caption{\label{fig:contact_density}Electron-positron contact density for $\rho_\text{Cl}=2.24$~a.u. 
For the hydrocarbons, plusses denote unenhanced values $\delta_{ep}^{(0)}$, while crosses denote enhanced values $\delta_{ep}$, as follows: 
blue, ethylene; green, propane; cyan, $n$-butane and isobutane; magenta, $n$-hexane.
For the chlorinated hydrocarbons, open symbols denote unenhanced values $\delta_{ep}^{(0)}$, while filled symbols denote enhanced values $\delta_{ep}$, as follows:
red circles, chlorinated methanes; indigo right-pointing triangles, chlorinated acetylenes; blue squares, chlorinated ethylenes; orange down-pointing triangles, chlorinated ethanes; green diamonds, chlorinated propanes; cyan left-pointing triangles, chlorinated butanes; magenta up-pointing triangles, $n$-hexyl chloride.
Black lines are the predictions of Eq.~(\ref{eq:cd_kappa_prop}): solid, $F=0.66$~a.u.; dashed, $F=0.20$~a.u.; dot dashed, $F=0.99$~a.u.
}
\end{figure}

The relationship between the contact density and the binding energy is analyzed more closely in Fig.~\ref{fig:contact_density}.
By plotting the contact density $\delta_{ep}^{(0)}$ or $\delta_{ep}$ against $\kappa=\sqrt{2\varepsilon_b}$, we see that it is indeed approximately proportional to $\kappa$; see Eq.~(\ref{eq:cd_kappa_prop}). One must keep in mind, however, that the linear dependence predicted by Eq.~(\ref{eq:cd_kappa_prop}) is expected to be valid only for nonpolar species. For polar molecules, the positron density near the molecule shows a stronger dependence on $\kappa $, as can be see, e.g., for the chlorinated methane family.

Analysing all the data in Fig.~\ref{fig:contact_density} and using linear regression, we find $F=0.20$~a.u. for the unenhanced contact density (with correlation coefficient 0.94), and $F=0.99$~a.u. for the enhanced contact density (with correlation coefficient 0.95); the corresponding lines are shown in Fig.~\ref{fig:contact_density}. The figure  also shows the line corresponding to $F=0.66$~a.u., which was found from an analysis of high-quality calculations of lifetimes of several positron-atom bound states \cite{Gribakin01}. 
We observe that the scaling factor $F$ is about 50\% greater for the molecular species studied here than for atoms. The reason for this is because in atoms, access of the positron to the electrons is impeded by the nuclear repulsion, while in molecules, the electrons are shared between the constituent atoms and can be accessed more easily by the positron \cite{Swann19}. 

Note also that previous calculations using the model-correlation-potential method for alkanes found $F=1.2$~a.u. \cite{Swann19}, which is 21\% larger than the value of $F=0.99$~a.u. found here. Indeed, looking at Fig.~\ref{fig:contact_density}, we see that the calculated values of $\delta_{ep}$ for $n$-butane, isobutane, and $n$-hexane are signficantly larger than the prediction of Eq.~(\ref{eq:cd_kappa_prop}) with $F=0.99$~a.u. 
A possible explanation for this is as follows.
For nonpolar or weakly polar molecules (like the alkanes), the positron wave function surrounds the entire molecule, so the positron can easily access electron density from any direction. On the other hand, for polar molecules (like many of the chlorinated hydrocarbons studied here), the positron is localized outside the negative end of the dipole, where it can only access electron density in a restricted region. Thus, if we have two molecules with the same positron binding energy, one nonpolar and one polar, we would expect the electron-positron contact density to be larger for the nonpolar molecule.
For example, the nonpolar dichloroacetylene and the polar methyl chloride ($\mu=2.31$~D) have approximately the same binding energy: $\varepsilon_b=24$~meV and 26~meV, respectively, for $\rho_\text{Cl}=2.24$~a.u. (see Table \ref{tab:binding_energies}). The corresponding values of the enhanced contact density are $\delta_{ep}=6.6\times10^{-3}$~a.u. and $4.2\times10^{-3}$~a.u. (see Table \ref{tab:contact_densities}): the contact density for dichloroacetylene is 1.5 times greater than for methyl chloride.

\section{\label{sec:conc}Conclusions}

Calculations and measurements of positron binding energies for a range of chlorinated hydrocarbons have been presented. 

The calculations were carried out using a model-potential method. The positron-molecule correlation potential accounts for long-range polarization of the molecule, while short-range correlations are parametrized by  cutoff radii whose values are specified for each type of atom within the molecule.
The trends in the calculated binding energies were discussed with regard to the molecular dipole moment, dipole polarizability, and geometry. For nonpolar or weakly polar molecules, the positron wave function surrounds the entire molecule, and binding energy is determined almost solely by the molecule's polarizability, with a larger polarizability leading to a larger binding energy. For more strongly polar molecules ($\mu\gtrsim1$~D), the positron is most likely to be found near the negative end of the dipole. Typically, this is where the one or more Cl atoms are located. The proximity of the positron to the highly polarizable Cl atoms makes the binding energy larger than that of a nonpolar molecule with the same total polarizability.
Also, for strongly polar molecules that have the same dipole moment and polarizability, the specific geometry of the molecule can play an important role in determining the binding energy. This was illustrated by $n$-propyl chloride and isopropyl chloride, which have very similar dipole moments and polarizabilities, yet the calculated binding energy for isopropyl chloride is 30\% larger than that for $n$-propyl chloride. This difference  results from the location of the Cl atom which marks the negative end of the molecular dipole. For isopropyl chloride, the Cl atom is bound to the central carbon and lies above the plane of the three C atoms, and the positron is close to all of them.
For $n$-propyl chloride, the Cl atom is at the end of the molecule and is coplanar with the three C atoms, so the positron interaction with the two more distant C atom is reduced.


Also presented here are new and reanalyzed experimental data for 14 molecules with an emphasis on hydrocarbons with various chlorine substitutions. The reanalysis of previous data corrects some inconsistencies, resulting in more accurate binding energy values suitable for the broad comparisons discussed here.


Overall, strong positive correlation between the calculated and measured binding energies is observed. In particular, theory predicts correctly the relative order of the binding energies for each chlorinated molecular family. At the quantitive level, larger discrepancies between theory and experiment are seen in the chlorinated ethylenes and $n$-hexyl chloride. For the chlorinated ethylenes, which are planar molecules, this may be due to anisotropy of the molecular polarizability, which is not accounted for in the present calculations. The reason for the discrepancy between theory and experiment for $n$-hexyl chloride is unclear.

{
Regarding the chloroethylene molecules, a recent paper \cite{Suzuki20} explored the positron binding to all of the six molecules discussed here using a density-functional theory (DFT) approach. The positron-molecule interaction was described using a correlation-polarization potential. It included a gradient correction with an adjustable parameter ($\beta $) whose magnitude was set by comparison with experimental binding energies. The authors of Ref.~\cite{Suzuki20} found that it was possible to choose a value of $\beta $ that could fit the measured binding energies for all three dichloroethylenes to within $\sim $10~meV. However, such a calculation strongly underestimated $\varepsilon _b^\text{exp}$ for tri- and tetrachloroethylene. For these molecules the best choice was effectively no gradient correction. This still led to large discrepancies between the binding energies obtained using different DFT approaches, the worst leading to a 30~meV disagreement with experiment. This again illustrates the difficulty experienced by theories in describing positron-molecule binding accurately.}

{
Besides the binding energies}, our calculations provided values of the electron-positron contact density in the bound state, which determines the lifetime of these states before annihilation. Broadly speaking, this quantity was found to be proportional to the square root of the binding energy, although it appears to have some dependence on the molecular dipole moment too. 
For two molecules, one nonpolar and one polar, with the same binding energy, the polar molecule is likely to have a smaller electron-positron contact density than the nonpolar molecule. This may be because for the polar molecule, the positron is localized near the negative end of the dipole, where it can only access electron density in a restricted region.

The main sources of uncertainty in the calculations are the choices for the cutoff radii and the lack of account for anisotropy of the molecular polarizability tensor.
If the cutoff radius for each type of atom can be chosen by reference to an accurate \textit{ab initio} calculation or an accurate measurement of the  binding energy for some reference data set, then using
these cutoff radii for other molecules should give reliable results; however, the lack of accurate \textit{ab initio} calculations of positron-molecule binding and the uncertainties present in the experimental data mean that a degree of uncertainty remains in the choices of cutoff radii for the model-correlation-potential method.
The lack of account for anisotropy of the polarizability tensor appears to be a significant problem for planar molecules. It may be possible to rectify this by using a different form for the model positron-molecule correlation potential that explicitly accounts for the anisotropy, e.g.,
\begin{align}
V_\text{cor}(\mathbf r) &= - \sum_{A=1}^{N_a} \sum_{i=1}^3 \frac{\alpha^A_{ii} \big[x_i-x_{Ai}\big]^2}{2 \lvert \mathbf r - \mathbf r_A \rvert^6}  \big( 1 - e^{-\lvert\mathbf r - \mathbf r_A \rvert^6/\rho_A^6} \big) ,
\end{align}
where the $x_i$ ($i=1$, 2, 3) are the Cartesian coordinates $x$, $y$, and $z$, with the corresponding axes chosen to be the principal axes \cite{polarizability_note}, and the $\alpha^A_{ii}$ are the components of the (diagonal) atomic hybrid polarizability tensor for atom $A$, with $\alpha_A = (\alpha^A_{xx} + \alpha^A_{yy} + \alpha^A_{zz})/3$ \cite{Miller90a}; cf.~Eq.~(\ref{eq:Vcor}). This will be a topic for future work.

Other areas for future investigation are calculations of binding energies for nitriles, alcohols, aldehydes, ketones, formates, and acetates, for which many binding energies are known from the measurements \cite{Danielson12}. Calculations can also be extended to the annihilation $\gamma$-ray spectra, for  which   the experimental data \cite{Iwata97} have only recently started to be investigated \cite{Green12,Ikabata18}. We have previously shown that the model-correlation-potential method can be used to calculate low-energy elastic scattering and direct annihilation rates for small nonpolar molecules \cite{Swann20}, and it may be possible to also use the method to compute direct annihilation rates for polar molecules.

\begin{acknowledgments}
This work has been supported by the  Engineering and Physical Sciences Research Council (EPSRC) UK, Grant No. EP/R006431/1 and the U.S. National Science Foundation, Grant No. PHY-2010699.
\end{acknowledgments}


\begin{thebibliography}{81}
\bibitem [{ann()}]{annrate_note}%
  \BibitemOpen
  \href@noop {} {}\bibinfo {note} {Equation (\ref{eq:ann_rate}) is valid when
  the total electron spin is zero, and the annihilation is into two $\gamma$
  rays.}\BibitemShut {Stop}%
\bibitem [{pol()}]{polarizability_note}%
  \BibitemOpen
  \href@noop {} {}\bibinfo {note} {By choosing the $x$, $y$, and $z$ axes to be
  the principal axes, the polarizability tensor $(\alpha_{ij})$ becomes
  diagonal. Then, the spherically averaged polarizability is given by
  $\alpha=(\alpha_{xx}+\alpha_{yy}+\alpha_{zz})/3$.}\BibitemShut {Stop}%
\end{thebibliography}

%

\end{document}